\documentclass{aa}
\usepackage[varg]{txfonts}
\usepackage{lscape}
\usepackage{graphicx}
\usepackage{colortbl}
\usepackage{amssymb}
\usepackage{amsmath}
\usepackage{natbib}
\bibpunct{(}{)}{;}{a}{}{,} 
\usepackage{times}
\usepackage{aas_macros}
\usepackage{threeparttable}
\usepackage{subfigure}
\usepackage{multirow}
\usepackage{float}

\usepackage{color}

\begin{document}

\title{Spatial distribution of stellar mass and star formation activity at 0.2<z<1.2 across and along the Main Sequence}
\titlerunning{Spatial distribution of SF activity in galaxies}
\authorrunning{L. Morselli}
\author{L. Morselli\inst{1}
  \and P. Popesso \inst{1} 
  \and A. Cibinel \inst{2}
  \and P. A. Oesch \inst{3}
  \and M. Montes \inst{4}
  \and H. Atek \inst{5}
  \and G. D. Illingworth \inst{6}
  \and B. Holden \inst{6}}

\institute{Excellence Cluster Universe, Boltzmannstr. 2, 85748 Garching bei M\"unchen, Germany
\and Astronomy Centre, Department of Physics and Astronomy, University of Sussex, Brighton, BN1 9QH, United Kindom
\and Observatoire de Gen\'eve, 51 Ch. des Maillettes, 1290 Versoix, Switzerland
\and School of Physics, University of New South Wales, NSW 2052, Australia
\and Institut d'astrophysique de Paris, CNRS UMR7095, Sorbonne Universite, 98bis Boulevard Arago, F-75014 Paris, France
\and UCO/Lick Observatory, University of California, Santa Cruz, CA 95064, USA}

\date{Received 2 November 2018 / Accepted 14 December 2018}


\label{firstpage}

\abstract{High-resolution multi-wavelength photometry is crucial to explore the spatial distribution of star formation in galaxies and understand how these evolve. To this aim, in this paper we exploit the deep, multi-wavelength Hubble Space Telescope (HST) data available in the central parts of the GOODS fields and study the distribution of star formation activity and mass in galaxies located at different positions with respect to the Main Sequence (MS) of star-forming galaxies. Our sample consists of galaxies with stellar mass $\geq 10^{9.5} M_{\odot}$ in the redshift range 0.2 $ \leq z \leq 1.2$. Exploiting 10-band photometry from the UV to the near-infrared at HST resolution, we derive spatially resolved maps of galaxies properties, such as stellar mass and star formation rate and specific star formation rate, with a resolution of $\sim 0.16$  arcsec. We find that the star formation activity is centrally enhanced in galaxies above the MS and centrally suppressed below the MS, with quiescent galaxies (1 dex below the MS) characterised by the highest suppression. The sSFR in the outer region does not show systematic trends of enhancement or suppression above or below the MS. The distribution of mass in MS galaxies indicates that bulges are growing when galaxies are still on the MS relation. Galaxies below the MS are more bulge-dominated with respect to MS counterparts at fixed stellar mass, while galaxies in the upper envelope are more extended and have S\'ersic indexes that are always smaller than or comparable to their MS counterparts. The suppression of star formation activity in the central region of galaxies below the MS hints at \textit{inside-out} quenching, as star formation is still ongoing in the outer regions.}

\keywords{galaxies: evolution - galaxies: star formation - galaxies: structure}

\maketitle
\section{Introduction}

Star forming (SFGs) and passive galaxies differ not only in their level of star formation activity, but in a large range of properties. SFGs dominate the low mass end of the galaxy mass function, while passive galaxies dominate the high mass end \citep[e.g.][]{2006ApJ...639L...1P,Pozzetti:2009gw,Drory:2009cp,2010ApJ...709..644I}. SFGs largely populate low density environments, while passive ones are mostly found in massive galaxy groups and clusters \citep[e.g.][]{1980ApJ...236..351D,2004ApJ...615L.101B,2005ApJ...629..143B,2009ARA&A..47..159B}. SFGs are mainly disk-dominated, while bulge-dominated structures are found to efficiently describe objects with little or no ongoing star formation activity \citep{1957PASP...69..291M, 1998ARA&A..36..189K,2003MNRAS.341...54K,2011ApJ...742...96W,2012ApJ...760..131C}. 

In the star formation rate (SFR) - stellar mass (M$_{\star}$) plane, SFGs and passive galaxies populate very different loci. SFGs occupy a very tight relation called the Main Sequence (MS), characterised by a slope ranging from 0.6 to 1.2 and a very small scatter of $\sim 0.2-0.3$ dex \citep[e.g.][]{2007ApJ...660L..47N,2007ApJ...670..156D,2007A&A...468...33E,2009ApJ...698L.116P,2010ApJ...721..193P,2010A&A...518L..25R,2011ApJ...730...61K,2012A&A...537A..58P,2012ApJ...754L..29W,2014MNRAS.442..509B,2014A&A...561A..86M,2014ApJS..214...15S,2015ApJ...801L..29R,2017MNRAS.468..849S,2017MNRAS.466.1192M}. Such a relation is already in place at z $\sim$ 4 \citep[e.g.][]{2014ApJS..214...15S} and is considered the dominant mode of star formation, leading to 90$\%$ of the stellar mass budget in the Universe today \citep{2015A&A...575A..74S}. In addition, its relatively small scatter suggests that star formation is regulated by secular evolution more than by stochastic events like mergers. 

The MS has been extensively studied to better understand the nature of its scatter, and thus reveal physical processes responsible, at a given galaxy stellar mass, of enhancing or suppressing star formation. Observational studies reveal that galaxy morphology changes across and along the MS. In particular, galaxies in the upper and lower envelop of the relation seem to be more bulge dominated than MS counterparts at fixed stellar mass, and the bulge prominence also increases with increasing stellar mass, when moving along the MS \citep{2011ApJ...742...96W,2014MNRAS.441..599B,2014ApJ...788...11L,Guo:2015kp,2017A&A...597A..97M}. \cite{2017A&A...597A..97M} conclude that, at any stellar mass, the MS is located at the minimum of the B/T distribution. Galaxies in the MS upper envelope are characterised by a blue, star forming bulge, while the lower envelope is occupied by compact structures with dead bulges.

From the theoretical point of view, \cite{2015MNRAS.450.2327Z} and Tacchella et al. (2016) propose that episodes of compaction and gravitational instabilities could drive gas towards the centre of galaxies. This would lead to the creation of a blue, star forming, central component. After the compaction event, gas consumption would then quench the central star formation region, leaving a quenched, red bulge. Subsequent episodes of compaction and quenching would move galaxies above and below the MS and along it and explain the observed scatter. It is thus clear that in order to understand the evolution of the star formation activity and thereby the migration of galaxies across the MS, it is mandatory to obtain observational data that can spatially resolve galactic structure \citep{2006MNRAS.371....2A,2014ApJ...785L..36A,2017A&A...597A..97M}. At low redshift, \cite{2017arXiv171005049S} find that galaxies classified as AGN/LI(N)ER typically have centrally-suppressed sSFR profiles, possibly indicating a relation between central quenching and AGN activity. They also find that suppression of sSFR, both central and at outer radii, happens in galaxies that are more bulge-dominated than typical star forming ones, further suggesting that the presence of a central spheroidal component might result in the cessation of star formation activity also in the disk, as suggested by \cite{2009ApJ...707..250M}. \cite{2017arXiv171008610L}, exploiting ALMA observations of three nearby MaNGA green valley galaxies, conclude that star formation activity has to be first suppressed in the bulge following the depletion of cold gas, and subsequently in the disk, thus implying that quenching has to be an \textit{inside-out} mechanism. Using MaNGA data, \cite{2017arXiv171100915E} find that galaxies above the MS are characterised by an enhancement in star formation activity that is more pronounced at small radii, while passive galaxies have suppressed star formation activity especially in the central region, thus confirming qualitatively the compaction/depletion scenario. \cite{2017arXiv171005034B} argue that a slow quenching process must result in the suppression of the star formation activity at all radii, and not only in the central regions, as they find that green valley galaxies have typically non star forming bulges and lower star formation in the disk. Other works, instead, find that nearby early type galaxies are characterised by slightly positive age gradients, thus favouring an \textit{outside-in} evolution of star formation in galaxies \citep[e.g.][]{2017MNRAS.465..688G}

At higher redshifts, \citet[hereafter N16]{2016ApJ...828...27N} use $H{\alpha}$ stacked profiles and find that SFR from $H{\alpha}$ above (below) the MS is enhanced (suppressed) at all radii, suggesting a scenario where the physical processes driving the change in star formation activity are independent of the host galaxy mass and act throughout the galactic disk in a coherent manner. \textit{Inside-out} quenching is instead advocated by \cite{2016ApJ...816...87M}, \cite{2015Sci...348..314T}, \cite{2016MNRAS.458..242T}, and \cite{2018ApJ...859...56T} to explain trends of sSFR, colours, and dust, observed in real and simulated galaxies. 

In this work, we exploit the combination of high-resolution optical-to-NIR photometric data from the GOODS+CANDELS surveys, with the deep UV observations collected with the WFC3/UVIS camera as part of the HDUV survey \citep{2018arXiv180601853O} to derive resolved stellar population maps. This allows us to study the spatial distributions of stellar mass and star formation activity as a function of the offset from the MS. In Sec.~\ref{sec:data} we describe the catalogues from which different quantities are taken (Sec.~\ref{sec:indicators}), the final sample used in this work (Sec.~\ref{sec:sample}), and the procedures to obtain the resolved maps of stellar properties (Sec.~\ref{sec:indicators}), the structural parameters of galaxies (Sec.~\ref{sec:galfit}), the radial profiles (Sec.~\ref{sec:profiles}) and the MS for the final sample (Sec.~\ref{sec:ms}). In Sec.~\ref{sec:results} we study the distribution of SFR, stellar mass, and sSFR, while in Sec.~\ref{sec:maps} we check the distribution of dust and UV luminosity. We discuss our results and their implications in Sec.~\ref{sec:discussion}.


\section{Data}
\label{sec:data}

In this work we use galaxies lying in the central part of the GOODS fields \citep{2004ApJ...600L..93G} covered by the Hubble Deep UV survey (HDUV, PI: P. Oesch), representing some of the deepest, high resolution UV data to date \footnote{The HDUV imaging data are available as high-level science products at \url{https://archive.stsci.edu/prepds/hduv/}.}. The HDUV is a legacy imaging program carried out with the WFC3/UVIS camera onboard HST that with a total of 132 orbits allows to get very deep images (27.5-28 mag) with the F275W and F336W filters, over a wide area of $\sim$100 arcmin$^2$ and with high spatial resolution ($<$0.1''). 

Available in the central parts of the GOODS fields are the photometric and spectroscopic observations taken as part of the GOODS, CANDELS, and 3DHST surveys with HST/ACS and WFC3 \citep{2004ApJ...600L..93G,2011ApJS..197...35G,2011ApJS..197...36K,2014ApJS..214...24S,2016ApJS..225...27M}, as well as the deepest far-IR data to date, obtained combining observations of the GOODS fields taken with PACS (Photodetector Array Camera and Spectrometer) as given by the combination pf the PACS Evolutionary probe survey \citep{2011A&A...532A..90L} and of the GOODS-Herschel survey \citep{2011A&A...533A.119E}, described in \citet[hereafter M13]{2013A&A...553A.132M}. The IR data allow an accurate determination of IR luminosity ($L_{IR}$), and thus of the obscured SFR of dusty high redshift star forming galaxies. Together with other available imaging data in the central parts of the GOODS fields, it is thus possible to construct SEDs over a large wavelength range, including the rest-frame UV for galaxies at z $>$ 0.2


\subsection{Redshifts, Stellar Masses, SFRs and structural parameters}
\label{sec:indicators}

Redshifts and stellar masses are taken from the public releases of 3D-HST data products \citep{2014ApJS..214...24S}. In particular, redshifts are ranked as spectroscopic, grism, and photometric \citep{2016ApJS..225...27M} and the best available estimate is used for the computation of the stellar masses \citep{2014ApJS..214...24S}, obtained from fitting the 0.3-8$\mu m$ SED range with the FAST code \citep{2009ApJ...705L..71K} using the \cite{2003MNRAS.344.1000B} stellar population synthesis models, a \cite{2003PASP..115..763C} IMF, solar metallicities, a \cite{2000ApJ...533..682C} dust attenuation law and a exponentially declining star formation history. 

SFRs are considered as the sum of the UV unobscured and the IR obscured contributions to the star formation activity of galaxies. For sources that are in the GOODS-Herschel catalogue of M13, SFR$_{IR}$ is computed from $L_{IR}$ using the Kennicutt relation \citep{1998ARA&A..36..189K}. The total $L_{IR}$, covering the range 8-1000$\mu m$, is estimated by fitting the flux densities at 70, 100 and 160 $\mu m$ (when available) with the SED templates of \cite{2011A&A...533A.119E}. For details on the procedure, see \cite{2013MNRAS.434.3089Z}.  For sources without counterparts in the GOODS-H catalogue (mainly galaxies with low SFRs, located in the lower envelope of the MS, and passive galaxies) SFR$_{IR}$ comes from the catalogue of \citet[hereafter W14]{2014ApJ...795..104W}. Briefly, W14 derived a photometric catalogue of Spitzer/MIPS 24$\mu m$ sources using the approach described in \cite{2014ApJS..214...24S}. The 24$\mu m$ flux density is used to estimate  $L_{IR}$ through a luminosity-independent conversion,  demonstrated to recover $L_{IR}$ in good agreement with Herschel-PACS estimates \citep{2011ApJ...742...96W}. The UV contribution to the SFRs is also taken from W14, and it is estimated from the 2800$\AA$ luminosity, following the work of \cite{2005ApJ...625...23B}. We underline here that for all the sources with $L_{IR}$ the SFR is given by SFR = SFR$_{IR}$ + SFR$_{UV}$. For galaxies with 1 $<$ S/N $<$ 3 at 24$\mu m$, W14 provides an upper limit of $L_{IR}$, and thus of SFR$_{IR}$. For sources undetected in 24$\mu m$ the SFR is purely based on the UV contribution, and thus it is a lower limit.


\subsection{Galaxy Sample}
\label{sec:sample}
 
The sample is mass selected and it has been built by selecting galaxies in the 3D-HST catalogue located within the HDUV footprints and satisfying the criteria: LogM$_{\star}>$ 9.5 M$_{\odot}$ and $0.2<z<1.2$. This is done in order to reach the high completeness level of $\sim$95$\%$ up to z$\sim$1, as shown in W14, \cite{2013ApJ...777...18M} and \cite{2013ApJ...769...31T}. As the aim of this work is to study the spatial distribution of star formation activity in galaxies, AGN contribution to the UV emission of a galaxy might represent a source of contamination. For this reason, we removed X-ray selected AGN by cross matching our sample with the X-ray selected AGN catalogue of \cite[for GOODS-N]{2010A&A...518L..26S} and \cite[][for GOODS-S]{2010ApJ...716..348B}. This results in the omission of $\sim5\%$ of the sources. In addition, visual inspection was carried out on the stellar mass maps of galaxies to avoid sources with corrupted stellar properties maps and merging systems. Following visual inspection,  The final sample is made of 712 galaxies, of which  $\sim65\%$ have spectroscopic redshifts and $\sim35\%$ have grism redshifts. In addition, 56$\%$ have a counterpart in M13, thus an estimate of $L_{IR}$ from Herschel/PACS. Of the remaining galaxies, 71$\%$ has S/N $>$ 1 at 24$\mu m$, 18$\%$ has $0<$S/N $<$ 1 at 24$\mu m$, and 11$\%$ has S/N $<$ 0 at 24$\mu m$.
In the following analysis, we split the sample in two redshift bins: [0.2:0.7] (\textit{\textit{low-z}} sample) and [0.7:1.2] (\textit{\textit{high-z}} sample).


\subsection{Resolved Stellar Population Maps}
\label{sec:maps}
In this paper, we exploit the multi-wavelength, high-resolution HST imaging data to obtain a resolved picture of galaxy evolution through stellar population maps. To build resolved maps, we follow the method described in \cite{2015ApJ...805..181C}. First, we match all the images in the different bands to the same resolution of the H-band, that with  $\sim0.16''$ is the filter with the worst PSF among the available ones. We then extract the pixel-by-pixel fluxes within an elliptical aperture equal to 1.5 times the Kron radius of each galaxy, and fit them with stellar population models. To overcome the problem of relatively low flux in individual pixels, we apply the ADAPTSMOOTH code of \cite{2009arXiv0911.4956Z} to the galaxy images in all the available filters. When, for a given pixel, the S/N is lower than a given threshold, the code replaces that value with an average of the values over a lager circular area. As explained in \cite{2015ApJ...805..181C}, an advantage of using ADAPTSMOOTH is the facility to smooth images in different filters on the same scale length, which is a necessary step to obtain self-consistent SEDs. We run the code on stacked images of all ACS and WCF3 stamps (to ensure that smoothing is applied on those pixels where the majority of the bands reach a low S/N while preserving the structural variations in the different filters) by requiring a minimum S/N = 5 and stop the algorithm when the averaging area reaches the maximum radius of 5 pixels. The smoothing pattern obtained as output is then applied to all available bands.

\begin{figure}
\centering
\includegraphics[scale=0.46]{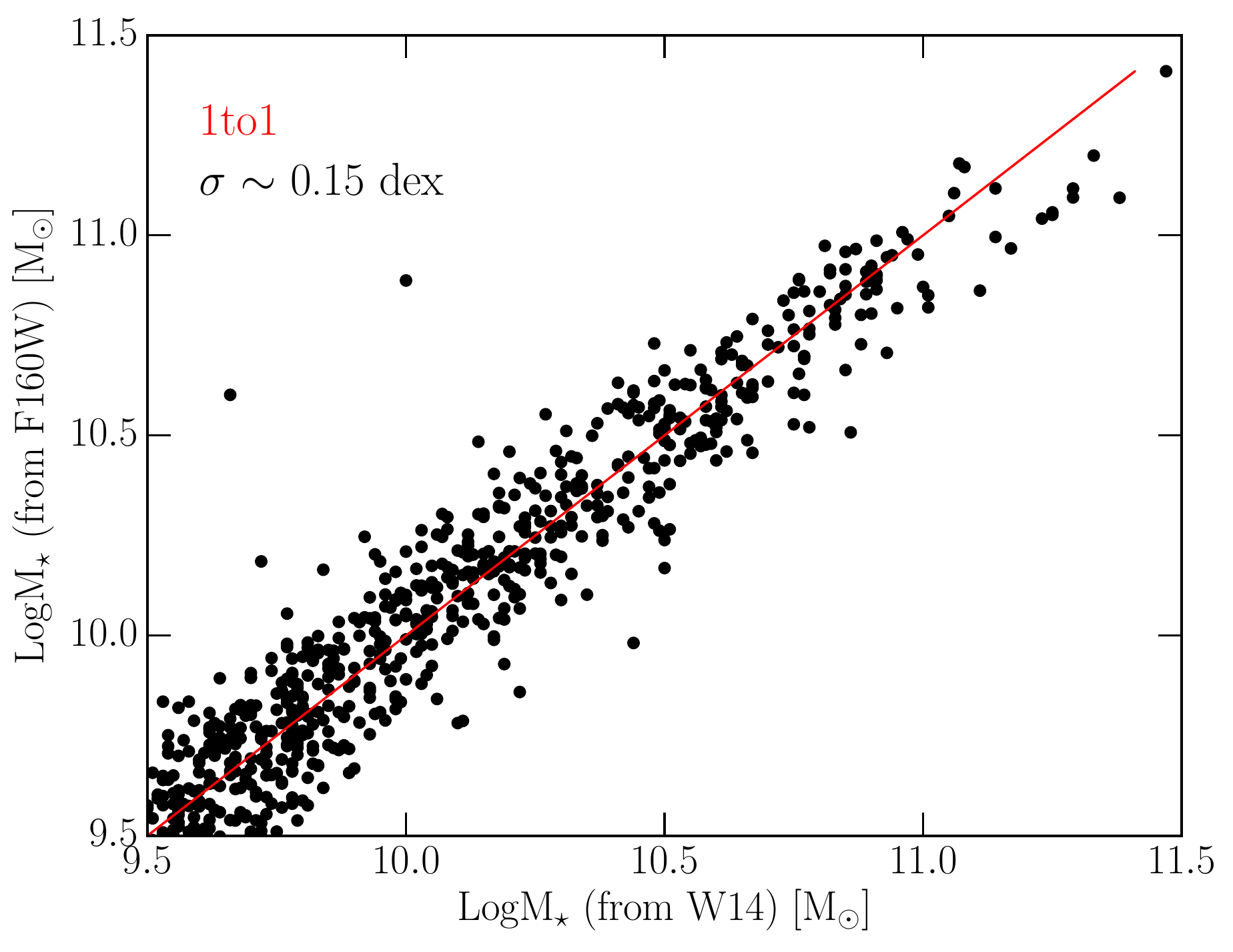}
\caption{Comparison between the stellar mass obtained from SED fitting in the work of W14, and the stellar mass computed as sum of the value in each pixel of the stellar mass maps created for this work, from the F160W images as explained in Sec.\ref{sec:maps} . The red line is the 1-to-1 relation. The total stellar mass derived from our resolved stellar mass maps is in very good agreement with the estimate obtained from the integrated photometry (scatter of 0.15dex). } 
\label{fig:mass_comp}
\end{figure}

\begin{figure}
\centering
\includegraphics[scale=0.46]{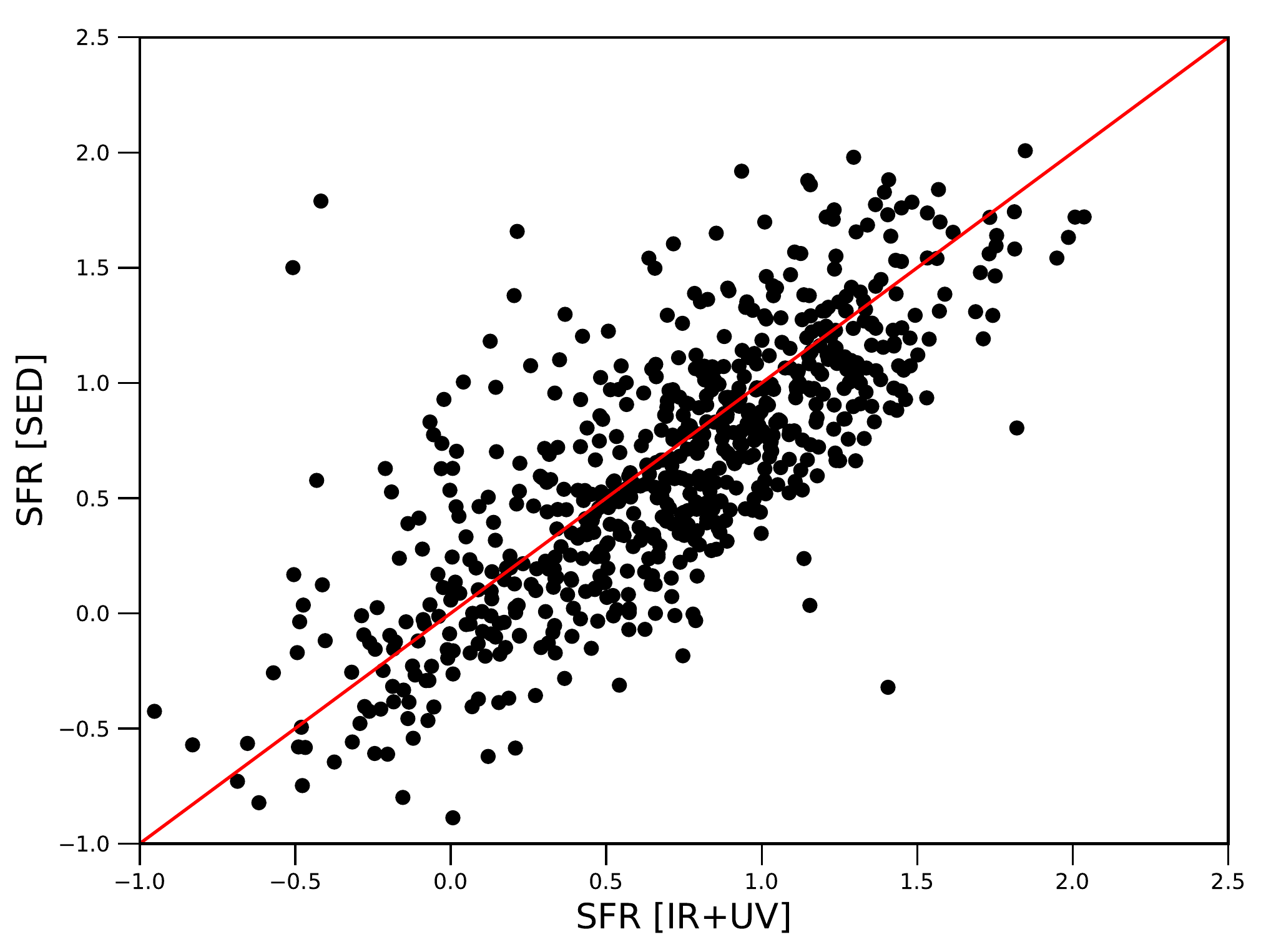}
\caption{Comparison between the SED-based integrated SFR, computed by summing the pixel-by-pixel values of SFR maps, and the UV+IR SFRs described in Sec.~\ref{sec:indicators}. } 
\label{fig:sfr_comparison}
\end{figure}

The pixel-by-pixel SED maps are then fitted with \textrm{LEPHARE} \citep{1999MNRAS.310..540A,2006A&A...457..841I} using the \cite{2003MNRAS.344.1000B} synthetic spectral library with a \cite{2003PASP..115..763C} IMF and a delayed exponential star formation history ($\psi \propto (t/\tau^2) \exp(-t/\tau)$). We apply the following constrains to the fits: 1) the typical timescale $\tau$ can vary between 0.01 and 10 Gyr in 22 steps, and 2) the template ages are selected between 100 Myr and the age of the Universe at the redshift of the source. We apply a \cite{2000ApJ...533..682C} extinction law, with E(B-V) varying between 0 and 0.9 mag and allowing for three values of metallicity Z (Z = Z$_{\odot}$, Z = 0.2 Z$_{\odot}$, Z = 0.4 Z$_{\odot}$). As result of the fitting procedure, we obtain median maps, where the value of each pixel is given by the median of the full probability distribution function from all the templates, and a best-fit map (i.e., minimum $\chi^2$). In the following analysis, we make use of the median maps. Fig.~\ref{fig:sfr_comparison} shows a comparison between the SFR obtained as sum over all pixels of the SED-based SFR maps and the SFR by UV + IR tracers (see Sec.~\ref{sec:indicators}). Despite a large scatter, there is good agreement between the two estimates of SFR.  

While for the SFR and E(B-V) we use the resolved maps as directly obtained from the SED-fitting, the derivation of the stellar mass maps requires a slightly different approach, motivated by the fact that we will use these maps to derive mass-based structural parameters with GALFIT.  In fact, as the maps obtained from SED fitting are defined within an aperture equal to the Kron radius and have zero values outside such radius, they do not include a sky component. This could lead to artifacts or unstable fitting with GALFIT. Therefore, we use the best-fit M/L maps obtained from SED fitting where we replace the zero values with the mean of the closest 15 pixels that have non-zero values. We then multiply these M/L maps with the $H$-band maps to get stellar mass maps that include the sky-noise contribution. To check the reliability of this procedure, in Fig.~\ref{fig:mass_comp} we show a comparison between the total stellar mass computed as sum of the values in each pixel of the mass map obtained from the $H$-band image, and the stellar mass in the 3D-HST catalogue. The two estimates are in good agreement, with a scatter of $\sim$0.15 dex. 

Finally, the UV luminosity maps are constructed from interpolation to the observed photometry, and they can be non-dust corrected, or corrected for dust using the UV beta slope - A$_{UV}$ relation \citep{1999ApJ...521...64M,2011ApJ...726L...7O}, where the UV beta slope is obtained from a linear fit to the best-fit SED model. 

\begin{figure*}
\centering
\includegraphics[width=0.95\textwidth]{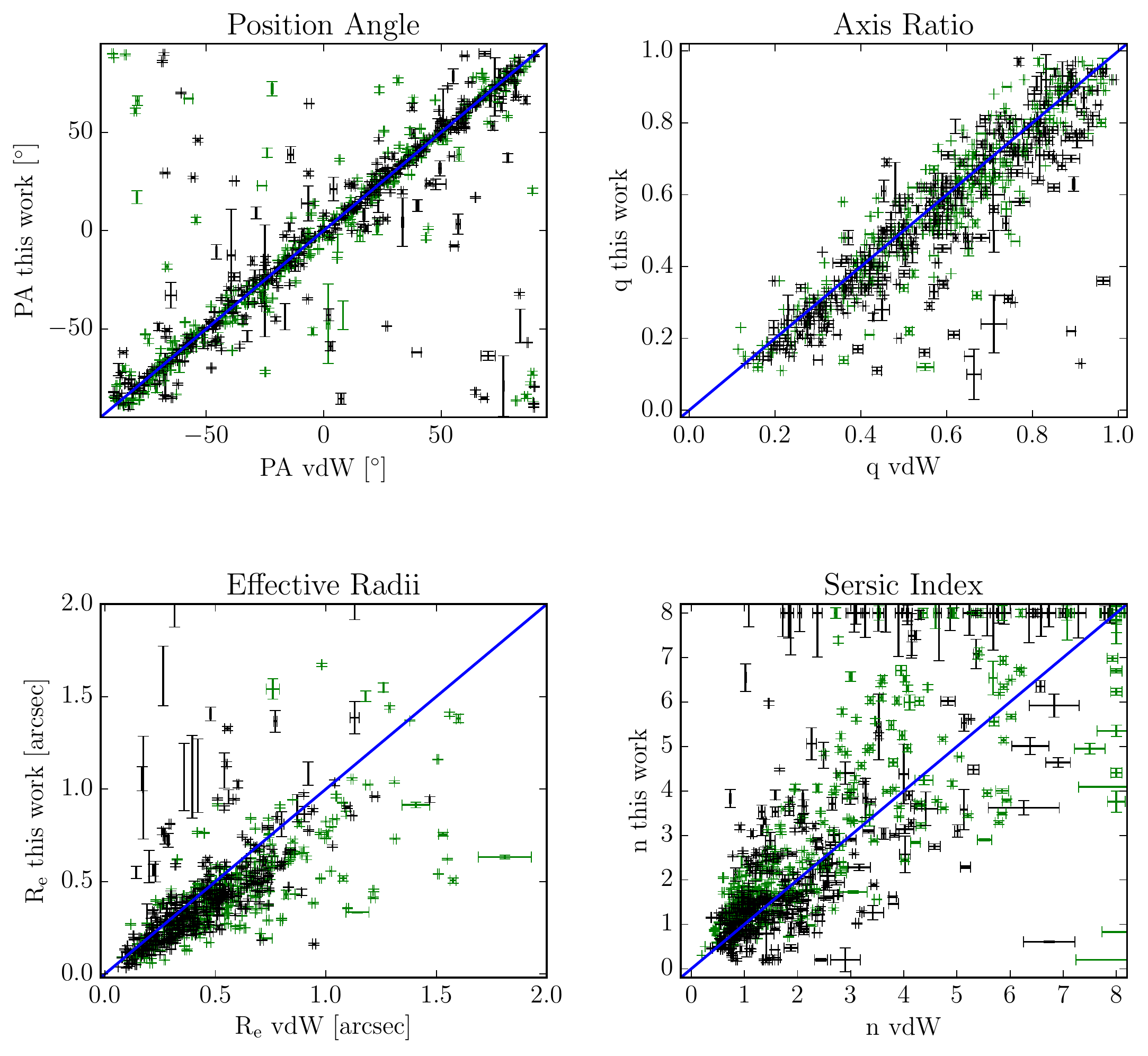}
\caption[Structural parameters in this work and vdW12]{Comparison between the mass-weighted structural parameters obtained from a single S\'ersic component GALFIT 2D fitting technique in this work and the light-weighted estimates of van der Wel et al. (2012). Top row: position angle (left) and axis ratio (right). Bottom row: effective radii (left) and S\'ersic index (right). In all four panels, the black symbols indicate galaxies with reliable fits in vdW12, while green ones mark galaxies with trustable fits in this work, but not in vdW12. The blue line is the 1-to-1 relation. For clarity, few points with extremely large errorbars have been omitted in this plot. The mass-weighted structural parameters obtained in this work are in good agreement with the light-weighted estimated of vdW12, especially for the axis ratio and the position angle. The majority of galaxies in the sample has mass-weighted effective radius smaller than the light-weighted value, and S\'ersic index larger than estimate of vdW12.  }
\label{fig:vdw-me}
\end{figure*}


\subsection{Mass-weighted structural parameters from GALFIT}
\label{sec:galfit}

Light-weighted structural properties of galaxies in GOODS-S and GOODS-N have already been studied as part of the work of \citet[][vdW12 hereafter]{2012ApJS..203...24V}. The authors use GALFIT \citep{2010AAS...21522909P} to derive structural parameters by fitting the cutouts of each galaxy in different filters (F160W, F125W and, when available, F105W) with a single S\'ersic profile. $55\%$ of the galaxies in our sample have, in vdW12, a flag corresponding to suspicious, bad, or non-existent fitting results. In this work, we compute mass-weighted structural parameters of galaxies, by fitting with GALFIT our 2D stellar mass maps using a single S\'ersic model.  Our fits are reliable for the $\sim94\%$ of the sample, while only $\sim6\%$ of fits retrieve unreliable estimates of the parameters (not within the input ranges) or non-existing fitting results. This difference most likely arises from the different input images that GALFIT needs to model. Here, in fact, GALFIT is run on the stellar mass maps obtained from pixel-by-pixel SED fitting of 10-bands photometric images that have been previously smoothed using ADAPSMOOTH (see Sec. 2.3 for the details on the creation of maps). Fig.~\ref{fig:vdw-me} shows a comparison of the mass-weighted structural parameters obtained in this work and the light-weighted ones obtained by vdW12 for the galaxies with reliable fitting results in our work. The panels show: the position angle ($PA$, top-left panel), the axis ratio ($q$, top-right panel), the effective radii (R$_\text{e}$, bottom-left panel) and the S\'ersic index ($n$, bottom-right panel), respectively. Galaxies that have reliable fits  both in this work and in vdW12 are marked in black; for these sources, light-weighted and mass-weighted structural parameters show a very good agreement. The green symbols mark galaxies with reliable fits in this work but not in vdW12. While the mass-weighted and light weighted estimates of the position angle and of the axis ratio of these galaxies show again a very good agreement, the scatter in the relation increases for the effective radius and the S\'ersic index. For these galaxies, the mass-weighted effective radius is often smaller than the light-weighted estimate, especially towards larger light-weighted R$_\text{e}$, in agreement with \cite{2013ApJ...763...73S}. The relation between the mass-weighted S\'ersic index and the light-weighted one is characterised by a large scatter, with the majority of galaxies having mass-weighted $n$ larger than the light-weighted estimate. Table \ref{table:n} is released with this paper\footnote{Table 1 is only available in electronic form at the CDS via anonymous ftp to cdsarc.u-strasbg.fr (130.79.128.5) or via http://cdsweb.u-strasbg.fr/cgi-bin/qcat?J/A+A/}, and contains the ID of the galaxy, its RA, DEC and redshift, and the mass-weighted structural parameters obtained from GALFIT: effective radius (in arcsec), Sersic index, axis ratio, and position angle.
 
\begin{figure*}
\centering
\includegraphics[scale=0.6]{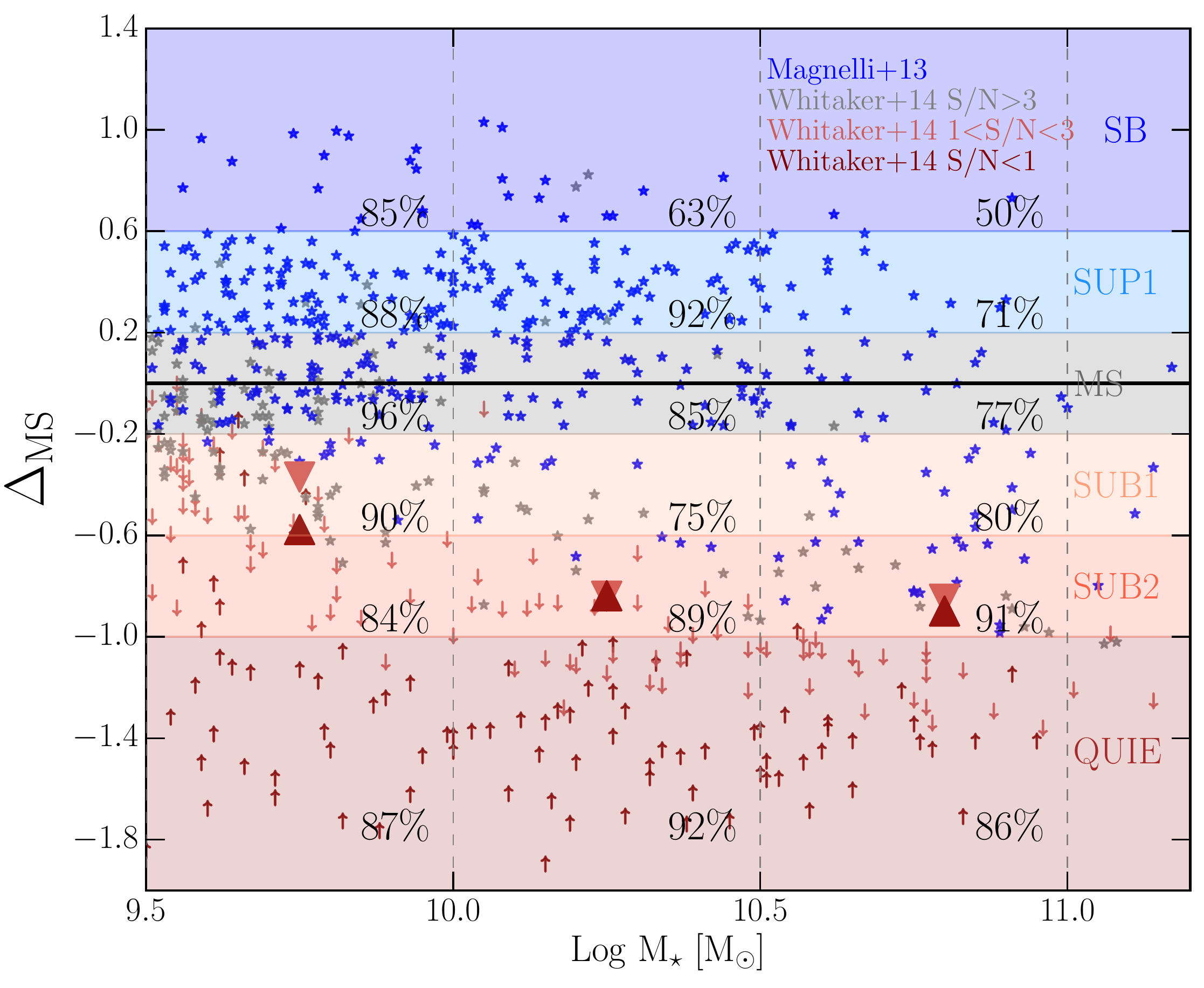}
\caption{Distribution of galaxies in the $\Delta_{MS}$ - LogM$_{\star}$ plane. Blue stars mark galaxies in the final sample with a counterpart the Herschel catalogue of M13. Grey stars indicate galaxies with no counterpart in M13, but detected with MIPS. Galaxies that are not detected in MIPS are shown with arrows: pink when the 24$\mu m$ is given as an upper limit, and maroon when the SFR comes from UV only. The triangles show the average distance from the MS of galaxies undetected in MIPS, for which the SFR from SED fitting was used to compute $\Delta_{MS}$. The percentages represent the completeness of the final sample (after removal of AGNs and corrupted data from visual inspection) with respect to the total HDUV sample in the same redshift bin. The shaded regions mark the bins defined according to the distance from the MS, and will be used from now on in this work.}
\label{fig:bins}
\end{figure*}

\begin{table*}
\centering
\caption{Example of the table released with this paper, containing the ID of the galaxy, its RA, DEC and redshift, and the mass-weighted structural parameters obtained from GALFIT: effective radius (in arcsec), Sersic index, axis ratio, and position angle.  }
\label{table:n}
\begin{tabular}{cccccccccccc} 
\hline
ID & RA & DEC & z & Re & err Re & n & err n & q & err q & PA & err PA \\
\hline
3138	&189.217644	&62.150245	&0.97	&0.19	&0.11	&2.85	&0.19	&0.78	&0.02	&47.47	&3.66\\
3164	&189.262815	&62.150191	&0.49	&0.12	&0.01	&1.73	&0.02	&0.29	&0.00	&-62.52	&0.14\\
3359	&189.273982	&62.151396	&1.16	&0.71	&0.25	&2.43	&0.06	&0.41	&0.00	&66.43	&0.41\\
3456	& 189.287288	& 62.152743	& 0.87	& 0.19     &0.06       &4.03	        & 0.16	& 0.65	& 0.01	&-2.76	&1.3 \\
3630	 & 189.232313	& 62.154862	& 0.47	& 0.38 	&0.04	&2.66	&0.02	&0.54	&0.0	        & -78.1	& 0.26\\
\hline
\end{tabular}
\end{table*}


 \subsection{Radial profiles}
\label{sec:profiles}

The 1D azimuthally averaged profiles of stellar mass, SFR, and sSFR of each galaxy are created from the 2D maps using the structural parameters obtained with GALFIT. First, a gaussian kernel is applied to the mass map to find the position of the peak, that we take as centre of the galaxy. Then, concentric ellipses are drawn on the 2D map of the analysed quantity (e.g. SFR or sSFR), where the major axis inclination depends on the galaxy position angle, together with the axis ratio used to define the ellipses. We build two types of radial profile: 1) the one normalised for the effective radii of each galaxy, where R$_\text{e}$ comes from the GALFIT fitting procedures; and 2) the physical one, where the radius is given in kpc. While the first method allows us to take care of the intrinsic distribution of R$_\text{e}$ with redshift (as galaxies at larger redshifts are more compact), it suffers from the large uncertainties in the estimate of R$_\text{e}$. On the other hand the second approach applied on a large redshift range, inevitably suffers from physical variation in the size of galaxies at different redshifts.

The median profile for a given mass and redshift bin is just the arithmetic median of all the regions between the same adjacent ellipses at a certain distance from the centre. We truncate the median profile at the radius that contains 50$\%$ of the galaxies of the given subsample.This is done to have a very high completeness in each region, in particular in the galaxy outskirts, that could be otherwise dominated by few particularly extended sources. In the specific case of the stellar mass, before computing the median each individual profile has been normalised  to enhance possible differences in the mass profile shape. The observed trends do not vary when considering the average profiles instead of the median ones. We checked the consistency of the profile procedure by comparing the mass obtained as sum of the values in each pixels of the map, and the total mass obtained from the integration of the radial profiles. As the two show very good agreement, we consider our method robust. 


\subsection{Main Sequence definition}
\label{sec:ms}

The aim of this work is to understand how the distribution of star formation activity and stellar mass change as a function of the location of a galaxy around the MS. To this aim,  we use the MS of \citet[][K11 hereafter]{2011ApJ...730...61K}, where SFRs are computed from 1.4GHz emission, considering a \cite{2003PASP..115..763C} IMF, \cite{2003MNRAS.344.1000B} models, and \cite{2000ApJ...533..682C} extinction curve. The K11 MS was chosen as it covers the redshift range $0.2<z<1.2$ in small redshift steps of 0.2 and has been computed using a relatively wide range in stellar masses ($\sim$2dex) that also includes low mass galaxies. While the former point is key in getting the right normalisation of the MS at each redshift, the latter is important to avoid possible biases due to the flattening of the MS. The MS flattening, still highly debated in literature \citep[e.g.][]{2014ApJ...795..104W,2016MNRAS.455.2839E}, if present would induce a bias in a selection of galaxies based on a MS computed only with massive sources. 

\begin{figure*}
\centering
\includegraphics[width=0.99\textwidth]{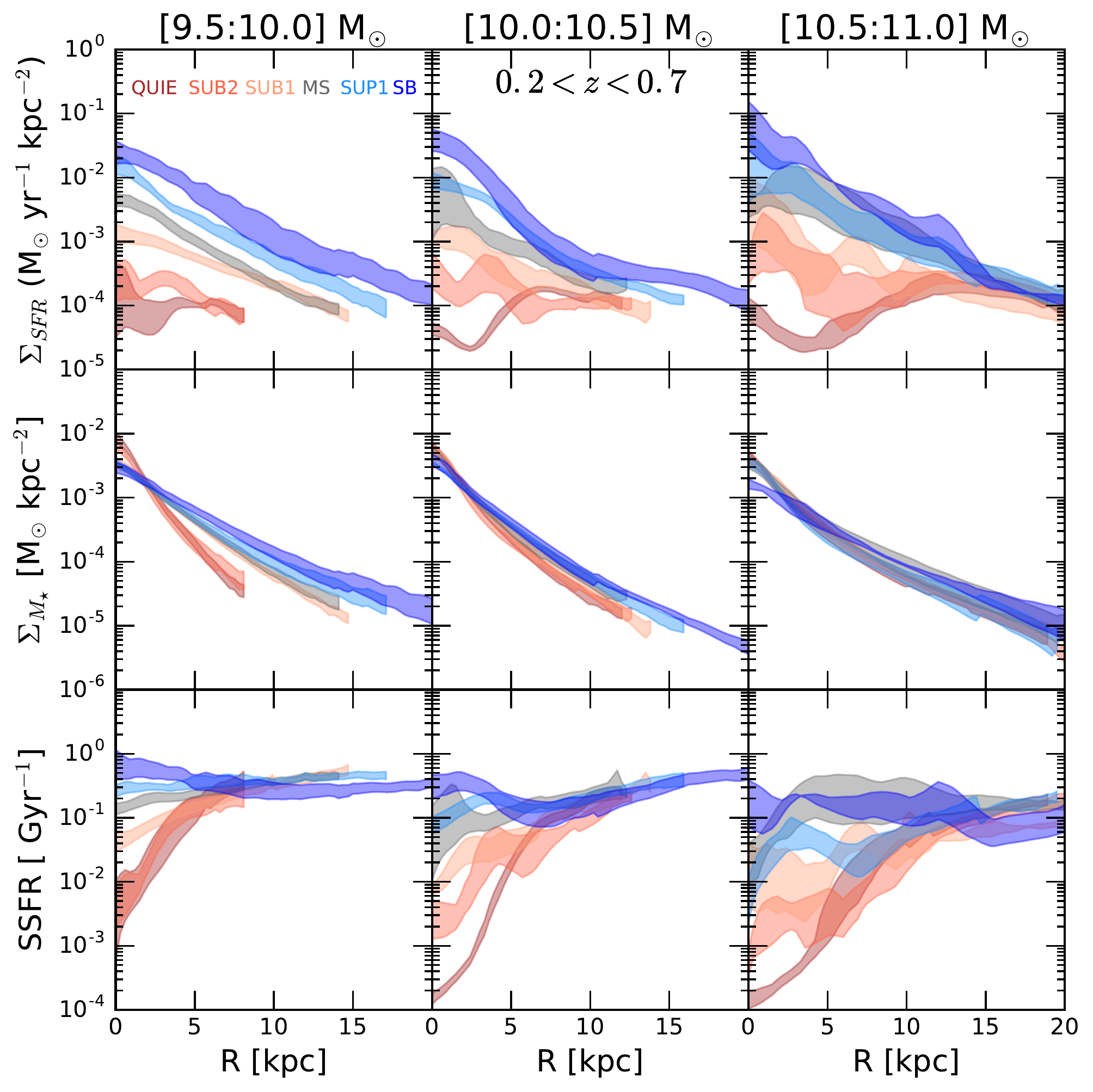}
\caption{Median SFR (first row), stellar mass (second row), and sSFR (third row) surface densities as a function of distance from the centre (in kpc) for galaxies in the redshift range [0.2:0.7]. The three columns indicate galaxies with different stellar masses: 9.5$<$Log(M$_{\star}$/M$_{\odot}$)$<$10.0 in the first, 10.0$<$Log(M$_{\star}$/M$_{\odot}$)$<$10.5 in the second, and 10.5$<$Log(M$_{\star}$/M$_{\odot}$)$<$11.0 in the third column, respectively. The thickness of the profiles in a given bin of distance from the centre is given by $\pm$ the standard deviation of the values in that bin, dived by the square root of the number of points in that bin. Different colours mark different bins of distance from the MS, as defined in Fig.~\ref{fig:bins}. The SSFR profiles of galaxies below the MS are characterised by a central suppression that becomes more significant with increasing distance from the MS.}
\label{fig:all_lowz}
\end{figure*}

Using the K11 definition of the MS at each redshift bin, we compute the distance of each galaxy from the relation at the galaxy's redshift. Fig.~\ref{fig:bins} shows the distribution of galaxies in the final sample on the $\Delta_{MS}$ - LogM$_{\star}$ plane, where $\Delta_{MS}$ = LogSFR$_{gal}$ - LogSFR$_{MS}$. The colours indicate different SFRs used to compute $\Delta_{MS}$, as explained in Sec.~\ref{sec:indicators}. Herschel detected galaxies (blue) mainly populate the upper envelope of the MS. MIPS-only detected galaxies (grey) mainly populate the lower envelope of the MS, especially at low stellar masses. MIPS upper limits with 1 $<$ S/N $<$ 3 (orange) populate the valley and quiescent region $\sim$ 1 dex below the MS. Galaxies undetected in MIPS (maroon), that only have the UV contribution to the SFR and thus are lower limits, are mainly found at distances larger that 1 dex below the MS. 
As a sanity check, we compute the average value of distance from the MS for galaxies that are undetected in MIPS, by summing the values in each pixel of the SED SFR maps. These values are marked in Fig.~\ref{fig:bins} with filled triangles. As expected, galaxies that are undetected in MIPS for which only the UV contribution to the SFR is considered are found closer to the MS when SFR from SED is used, but still well below the relation. 

We define six subsamples of galaxies based on their distance from the MS, chosen to have enough statistics and, at the same time, to take into account the average scatter of the MS ($\sim$0.2 - 0.3 dex). All galaxies with $\Delta_{MS}>0.6$ constitute the starburst (SB) sample, indicated with the blue shaded region. Galaxies in the upper envelope of the MS, with $0.2<\Delta_{MS}<0.6$, make up the SUP1 sample (light blue area in Fig.~\ref{fig:bins}). The MS sample is comprised of galaxies that have $-0.2<\Delta_{MS}<0.2$ (gray area in Fig.~\ref{fig:bins}). The lower envelope on the MS is divided in the SUB1 and SUB 2 samples, that include galaxies with $-0.6<\Delta_{MS}<-0.2$ (light pink area in Fig.~\ref{fig:bins}), and $-1.0<\Delta_{MS}<-0.6$ (orange area in Fig.~\ref{fig:bins}), respectively. Finally,  all galaxies with $\Delta_{MS}<-1.0$ form the quiescent (QUIE) sample marked by the shaded red colour. We note that due to the limited area covered by the HDUV survey, the number of real starbursts (usually defined in literature as galaxies with $\Delta_{MS}>1.0$) is relatively small. Tab.~\ref{table:n} shows the number of galaxies in each bin of mass, redshift, and distance from the MS. The completeness of our final sample in each stellar mass and distance from the MS bin is always larger than $50\%$, as indicated by the numbers in Fig.~\ref{fig:bins}.

\begin{table*}
	\centering
	\caption{Number of galaxies in each bin of distance from the MS, stellar mass, and redshift.}
	\label{table:n}
	\begin{tabular}{rlccc} 
		\hline
		        &     & Log(M$_{\star}$/M$_{\odot}$) & Log(M$_{\star}$/M$_{\odot}$) & Log(M$_{\star}$/M$_{\odot}$) \\
		$\Delta_{MS}$ & Sample & 9.5:10.0 & 10.0:10.5 & 10.5:11.0\\
		\hline
		&   & \textit{high-z} \ \ \textit{low-z}  & \textit{high-z}  \ \ \textit{low-z}  & \textit{high-z}  \ \ \textit{low-z}\\ 
		$\Delta_{MS} < -1$ & QUIE & 21\qquad 8 & 29\qquad 20 & 34\qquad 9\\
		$-1\le \Delta_{MS} < -0.6$ & SUB2 & 15\qquad 6 & 17\qquad 8 & 24\qquad 7\\
		$-0.6\le \Delta_{MS} < -0.2$ & SUB1 & 49\qquad 16 & 10\qquad 5 & 12\qquad 4\\
		$-0.2\le \Delta_{MS} < 0.2$ & MS & 102\qquad 28 & 31\qquad 14 & 17\qquad 7\\
		$0.2\le \Delta_{MS} < 0.6$ & SUP1 & 50\qquad 40 & 43\qquad 22 & 11\qquad 6\\
		$\Delta_{MS} \ge 0.6$ & SB & 5\qquad 12 & 10\qquad 7 & - \qquad  2\\
		\hline
	\end{tabular}
\end{table*}


\section{Results}

\subsection{SFR, M$_{\star}$, and sSFR profiles}
\label{sec:results}

In Fig.~\ref{fig:all_lowz}, we show the median SFR (first row), M$_{\star}$ (second row), and sSFR (third row) profiles in the different bins of distance from the MS and stellar mass for the \textit{\textit{low-z}} subsample. The profiles are computed as a function of the physical distance from the centre in kpc. The thickness of the profiles in a given bin of distance from the centre is given by $\pm$ the standard deviation of the values in that bin, dived by the square root of the number of points in that bin.The three columns of Fig.~\ref{fig:all_lowz} represent three stellar mass bins: 9.5 $<$ Log(M$_{\star}$/M$_{\odot}$) $<$ 10.0, 10.0 $<$ Log(M$_{\star}$/M$_{\odot}$) $<$ 10.5, and 10.5 $<$ Log(M$_{\star}$/M$_{\odot}$) $<$ 11.0, respectively. In all panels, the colours correspond to galaxies at different distances from the MS: from passive (red) to MS galaxies (grey) and upper MS galaxies (blue), as defined in Fig.~\ref{fig:bins}.   
For galaxies with $9.5<$ Log(M$_{\star}$/M$_{\odot}$) $<10$, the SFR profile of galaxies well below the MS (QUIE and SUB2) is almost flat at every radii, while the SFR profiles of galaxies around the MS or above it increases towards the inner radii. For larger stellar masses, the increase of SFR in the central part is more enhanced for galaxies at higher SFR, with SB showing the largest SFR surface density in the inner $\sim$ 5 kpc. At R $<$ 5 kpc, the difference between the SFR profiles of QUIE and SB galaxies with Log(M$_{\star}$/M$_{\odot}$) $>$ 10 reaches 3 orders of magnitudes. At larger radii (r $>$ 5 kpc) the SFR surface density decreases significantly in galaxies above and on the MS, reaching values comparable with the ones of galaxies well below the relation, that on contrary are characterised by a dip in the SFR distribution at small radii.

\begin{figure*}
\centering
\includegraphics[width=0.99\textwidth]{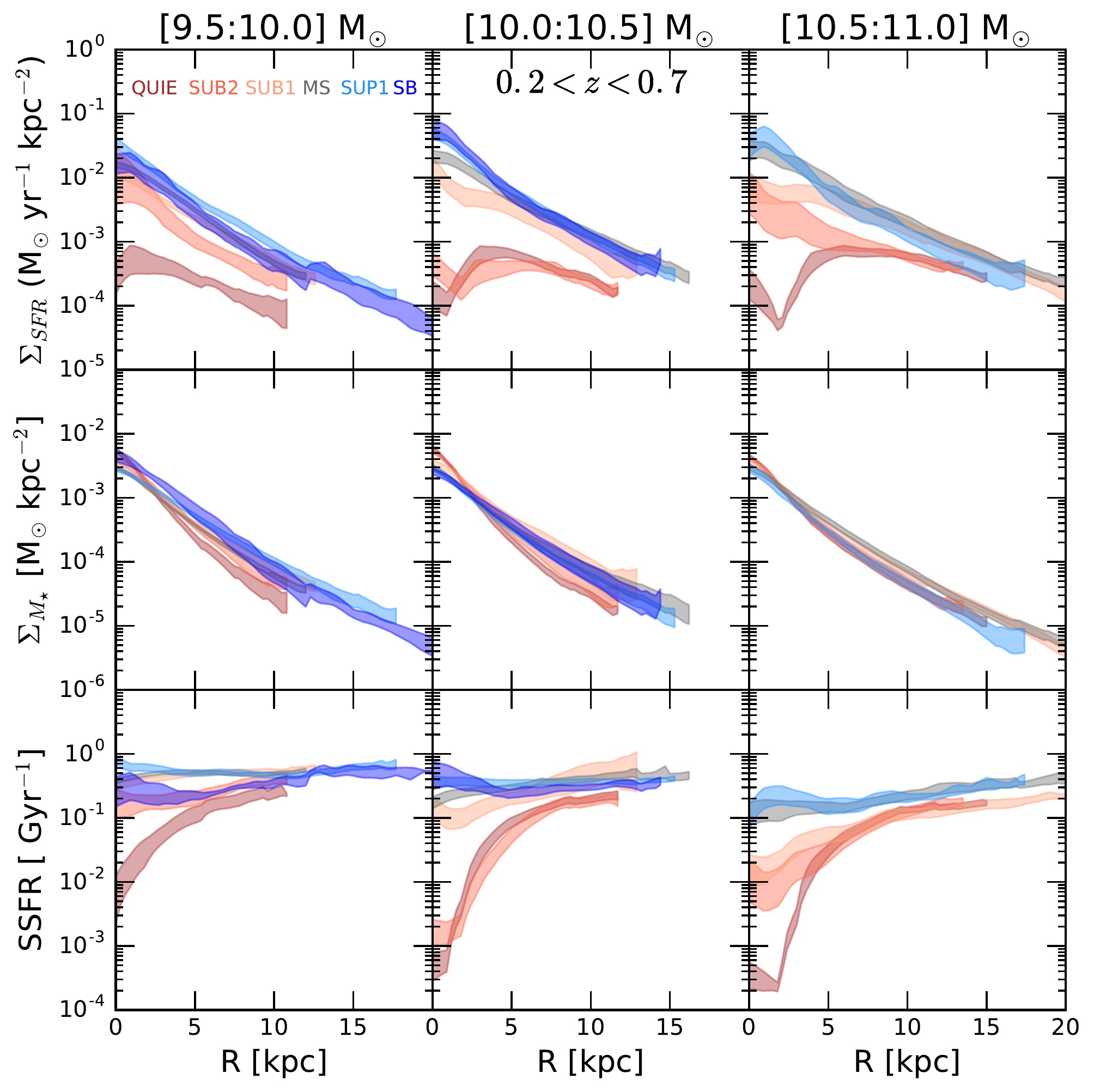}
\caption{Same as Fig.\ref{fig:all_lowz}, but for galaxies in the redshift range [0.7:1.2]. Also in the \textit{high-z} sample we observe that the sSFR of galaxies below the MS is significantly lower in the central regions with respect to MS counterparts.}
\label{fig:high_z}
\end{figure*}

The second row of Fig.~\ref{fig:all_lowz} shows the stellar mass profiles of galaxies. We recall here that each single profile has been normalised for the total stellar mass obtained integrating the profile, to account for differences in M$_{\star}$ within the 0.5 dex bin, before the computation of the median profile. As expected, passive galaxies are always less extended and more concentrated with respect to MS galaxies. MS galaxies become more extended with increasing stellar mass and their mass distribution seems to become more centrally concentrated, consistent with the idea of the build-up of the bulge component. Galaxies located above the MS are more extended than MS counterparts, and show no central enhancement of their stellar mass distribution. 

The third row of Fig.~\ref{fig:all_lowz} shows the sSFR profiles of galaxies. At large radii, r $>$ 5 kpc, the median sSFR profile does not show significant differences for the subsamples of galaxies located at various distances from the MS; galaxies with the largest SFRs and passive ones have the same sSFR surface density, or they differ of $\lesssim$ 0.5 dex. At small radii, r $<$ 5 kpc, the sSFR profiles depend dramatically on the distance from the MS. The sSFR profiles of galaxies below the MS are characterised by a decrease which is proportional to the distance from the relation, with quiescent galaxies showing the largest decline. MS and SUP1 galaxies are characterised by a small decline of sSFR in the inner radii, while SB galaxies have flat or enhanced sSFR profile in the innermost 5 kpc. 
 

Figures \ref{fig:high_z} shows the median SFR,  stellar mass, and sSFR profiles of galaxies in the \textit{high-z} sample, computed as a function of the physical distance from the centre in kpc. No significant evolution within the two redshift bin is observed, as the trends in the \textit{high-z} sample are very similar to the \textit{low-z} one. The only difference concerns the SFR profile of SB galaxies (notice that we do not have SB galaxies in the most massive bin). While in the \textit{low-z} sample the SB SFR profile is the one characterised by the largest SFR surface density at small radii, in the \textit{high-z} one the SFR profile of SB galaxies is consistent with the one of SUB1 and MS galaxies. The trends observed in Fig.~\ref{fig:all_lowz} and Fig.~\ref{fig:high_z}, in which the surface densities are shown as a function of the physical distance from the centre in kpc, are also observed when the surface densities are computed as a function of the distance normalised by the galaxies' effective radii (see Appendix \ref{app}). 

\begin{figure}
\centering
\includegraphics[width=0.43\textwidth]{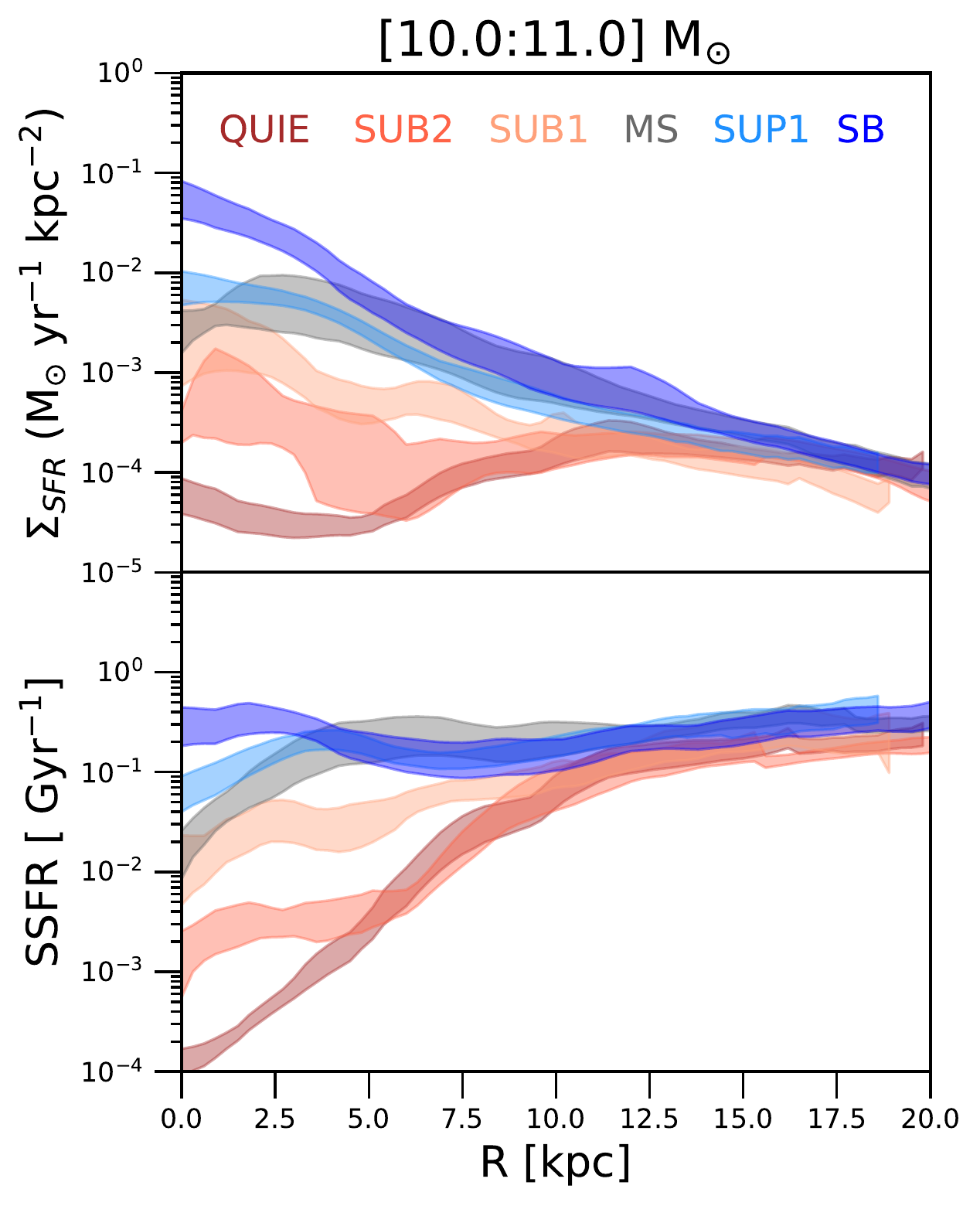}
\caption{Median SFR (first row) and sSFR (second row) surface densities as a function of distance from the centre (in kpc) for all the galaxies in the redshift range [0.2:0.7] and stellar mass range 10.0$<$Log(M$_{\star}$/M$_{\odot}$)$<$11.0 that have an effective radius larger than 0.45 arcsec. The observed trends are similar to the ones of the whole galaxy sample, independent on the galaxy size.}
\label{fig:ext}
\end{figure}

As explained in Section \ref{sec:maps}, the images in the different bands are PSF-matched to the F160W image prior to the pixel-by-pixel SED fitting routine. To investigate possible contamination due to PSF effects in the trends observed in the radial profiles, we selected the most extended galaxies in the low redshift range, for which PSF-related effects should be minimised. In Fig.~\ref{fig:ext} we show the radial profiles of galaxies that have an effective radii larger than 3 times the PSF FWHM in the H band, thus $R_e>0.45''$. The sample inevitably suffers of small statistics, thus we consider all galaxies in the stellar mass range  10 $<$ Log(M$_{\star}$/M$_{\odot}$) $<$ 11.0. The trends observed in Figures ~\ref{fig:all_lowz} and ~\ref{fig:high_z} characterise also the SFR and SSFR profiles of the most extended galaxies in our lower redshift sample, thus proving that our analysis is robust against PSF-related effects.

\begin{figure*}
\centering
\includegraphics[width=0.99\textwidth]{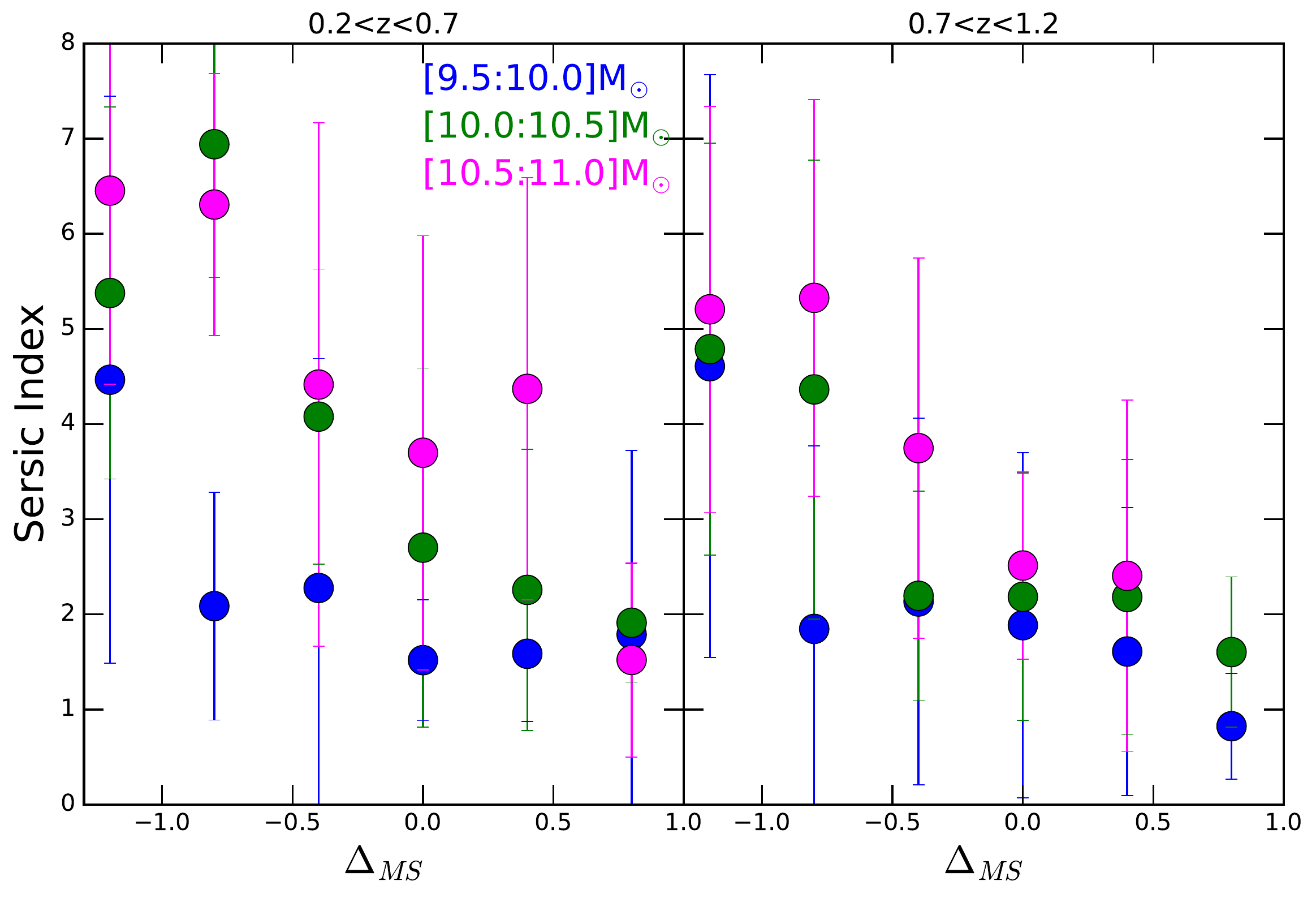}
\caption{Mass-weighted S\'ersic index, obtained from GALFIT, as a function of distance from the MS, $\Delta_{MS}$, for the low and high redshift samples (left and right panels, respectively). The three stellar mass bins are represented with different colours. At all stellar masses, galaxies below the MS are more compact than MS counterparts, while galaxies above the MS have similar or lower S\'ersic index than MS sources. An increase of the S\'ersic index is visible on the MS relation, with increasing stellar mass.}
\label{fig:sersic_mine}
\end{figure*}

To further study the mass distribution in galaxies above/below the MS, both in the low and high redshift sample, we show in Fig.\ref{fig:sersic_mine} the distribution of the average S\'ersic index $n$ as a function of the distance from the MS, $\Delta_{MS}$. The S\'ersic index used here is obtained as output of the GALFIT fitting procedure, as explained in Sec.~\ref{sec:galfit}. The errorbars in this analysis are computed via bootstrapping. We observe, as discussed qualitatively when analysing the stellar mass profiles, a quasi-monotonic decrease of $n$ with increasing $\Delta_{MS}$ in each stellar mass and redshift bin: galaxies below the MS are more concentrated than galaxies located on the MS or above it. Galaxies on the MS become more concentrated with increasing stellar mass, transforming from pure disks structures to bulge+disk ones. Also for galaxies below the MS the $n$ increases with increasing stellar mass. No increase in $n$ is observed in galaxies above the MS with respect to MS counterparts at fixed stellar mass, as found for the local counterparts in Morselli et al. (2017). 

\begin{figure*}
\centering
\includegraphics[scale=0.55, angle=270]{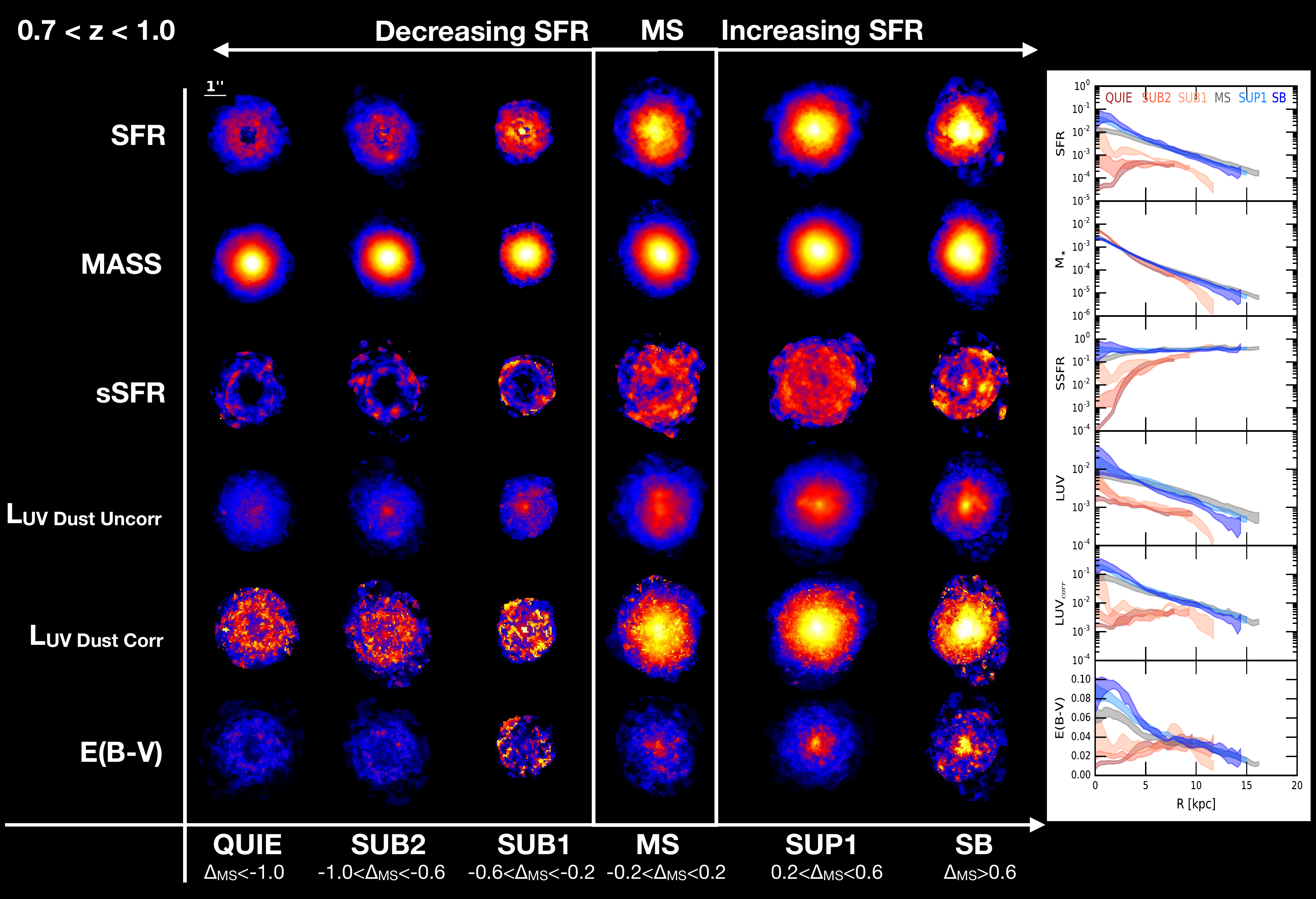}
\caption[Stacked SFR, M$\star$, sSFR, L$_{UV}$ and E(B-V) maps ]{From top to bottom: stacked maps of SFR, stellar mass, sSFR, uncorrected L$_{UV}$, dust corrected L$_{UV}$, and E(B-V) of galaxies with axis ratio $q>0.5$, $0.05''<re<0.8''$, $0.7<z<1.0$, and $10.0<$Log(M$_{\star}$/M$_{\odot}$)$<10.5$. The six columns are the six bins of distance from the MS. On the right side of the figure, the median radial profiles of galaxies in each bin are showed using the same color code of Fig. 4 and Fig. 5. The trends shown in this figure hint towards inside-out quenching below the MS and enhanced SF activity above the MS. }
\label{fig:maps}
\end{figure*}


\subsection{Stacked maps}
\label{sec:maps}

From what has been discussed thus far, the star formation activity in the central part of a galaxy seems to strongly correlate with the location of the galaxy in the LogSFR-LogM$_{\star}$ plane. Nevertheless, the sSFR and SFR surface density profiles from which these conclusions are drawn come from SED fitting, which is very sensitive to dust and, in general is the least stable method to derive SFRs. For this reason, we decided to further investigate the distribution of dust and UV luminosity (corrected and not-corrected for dust). Fig.~\ref{fig:maps} shows the mean stacked maps of SFR, stellar mass, sSFR, not-corrected UV luminosity, dust-corrected UV luminosity and E(B-V). The stacked maps have been built by excluding galaxies with axis ratio (obtained with GALFIT) $q<0.5$ and $0.05''<re<0.8''$. These cuts avoid biases due to highly inclined galaxies, or to extremely compact/extended galaxies that are not representative of the galaxy population in the bins. We focus on the range $0.7<z<1.0$ and $10<$ Log(M$_{\star}$/M$_{\odot}$)$ <10.5$ so as to ensure the best statistics while trying to limit the width of the redshift range, as intrinsic differences in the shape of galaxies at different cosmic times can bias the results, as the individual maps are not normalised to the effective radii before stacking. In addition, the radial profiles computed as explained in Sec. 2.5 are also shown in Fig.~\ref{fig:maps} to better visualise trends (with the same color-coding used in Fig. 4 and Fig. 5). 

The SFR maps show a monotonic increase of the SFR in the centre of galaxies when moving towards larger integrated SFRs. We underline here that even if the star formation activity is taking place in clumps in individual galaxies, the stacking procedure produces a "smooth" distribution of SFR in the disk. In the SB bin, the distribution of SFR (as well as other properties) looks more disturbed, as the number of galaxies in this bin is small. Nevertheless, the increase of SFR in the central part is visible. The mass distribution shows that galaxies below the MS, in the quiescent region of the LogSFR-LogM$_{\star}$ plane, are less extended and more concentrated than MS and upper MS galaxies, reflecting the trends in S\'ersic index observed both in the \textit{low-z} and \textit{high-z} sample (Fig.\ref{fig:sersic_mine}). As discussed above, galaxies above the MS are more extended than MS counterparts. sSFR maps enhance the trends visible in the SFR maps. Below the MS we can clearly see that star formation activity is taking place only in the outskirts of galaxies. The position of the clumps of star formation obtained as a result of stacking analysis are irrelevant to this study. However, note that the "hole" (or the lack of star formation) seen in the central part is a common feature of all the galaxies below the MS. The sSFR progressively increases in the central part of  galaxies from the quiescent bins towards the MS and above it, but it is interesting to observe that also on the MS the sSFR in the outskirts is larger than in the inner region of galaxies. Above the MS, the distribution is smoother in the SUP1 sample. In the SB sample, the distribution reveals an increase of sSFR in the central part of the galaxy, as well as other clumps of SF in the outskirts. The UV luminosity maps also show a trend of increasing L$_{UV}$ in the centre with increasing integrated SFR, even without dust correction. The distribution is "flat" for QUIE galaxies and progressively peaks in the central part of galaxies when moving towards larger integrated SFRs. Correcting for dust reveals a clumpy distribution of UV luminosity in the outskirts of galaxies in the QUIE and SUB2 samples, and further unveils the presence of a central depression of UV luminosity not visible when dust correction is not taken into account. When dust correction is applied, the UV luminosity in the central part of SB galaxies becomes more peaked than for SUP1 and MS galaxies. Finally, the trends seen in dust distribution are analogous to the ones seen for the SFR. A central hole in $E(B-V)$ is visible in the stacked maps of QUIE and SUB2 galaxies, while for the remaining bins the distribution peaks at the centre, and the peak value increases with increasing integrated SFR. 
We underline here that the average trends visible in the stacked maps of Fig.~\ref{fig:maps} are not biased by our stacking procedure, or by selection effects caused by limiting the sample to nearly face-on galaxies. The visual inspection of all maps confirms that, while the extension of the central region might vary from system to system, the central "hole" in sSFR in galaxies below the MS and the central SF enhancement in galaxies above it are visible in all systems of the two categories, respectively.  

The analysis of the stacked maps of galaxy properties presented in Fig.~\ref{fig:maps} shows that the radial trends observed in SFR and sSFR are reliable against dust correction and thus underline the importance of the central region of galaxies in determining the scatter of the MS and in understanding the quenching mechanism. We will discuss the possible scenarios arising from such trends in the next section.


\section{Discussion}
\label{sec:discussion}

We discuss here how our results compare with others in literature, as well as the emerging picture on how galaxies move in the LogSFR-LogM$_{\star}$ plane.


\subsection{SFR and sSFR profiles}

The SFR and sSFR profiles of galaxies suggest a fundamental role of the star formation activity in the central region in determining the location of a source around the MS and, thus, in the LogSFR-LogM$_{\star}$ plane. 

At fixed stellar mass, galaxies with the highest (lowest) SFR exhibit also the highest (lowest) SFR surface density within the inner 5 kpc. This leads to a clear trend as a function of the distance from the MS, with centrally active galaxies populating the region of the sequence and its upper envelope, and centrally suppresses galaxies located below the relation. We do observe a difference in the SFR level of the outskirts. However, this variation, in particular in the sSFR profile, is much lower (0.5-0.7dex) with respect to the variation in the central region (2-3 dex). 
\begin{figure}
\centering
\includegraphics[width=0.4\textwidth]{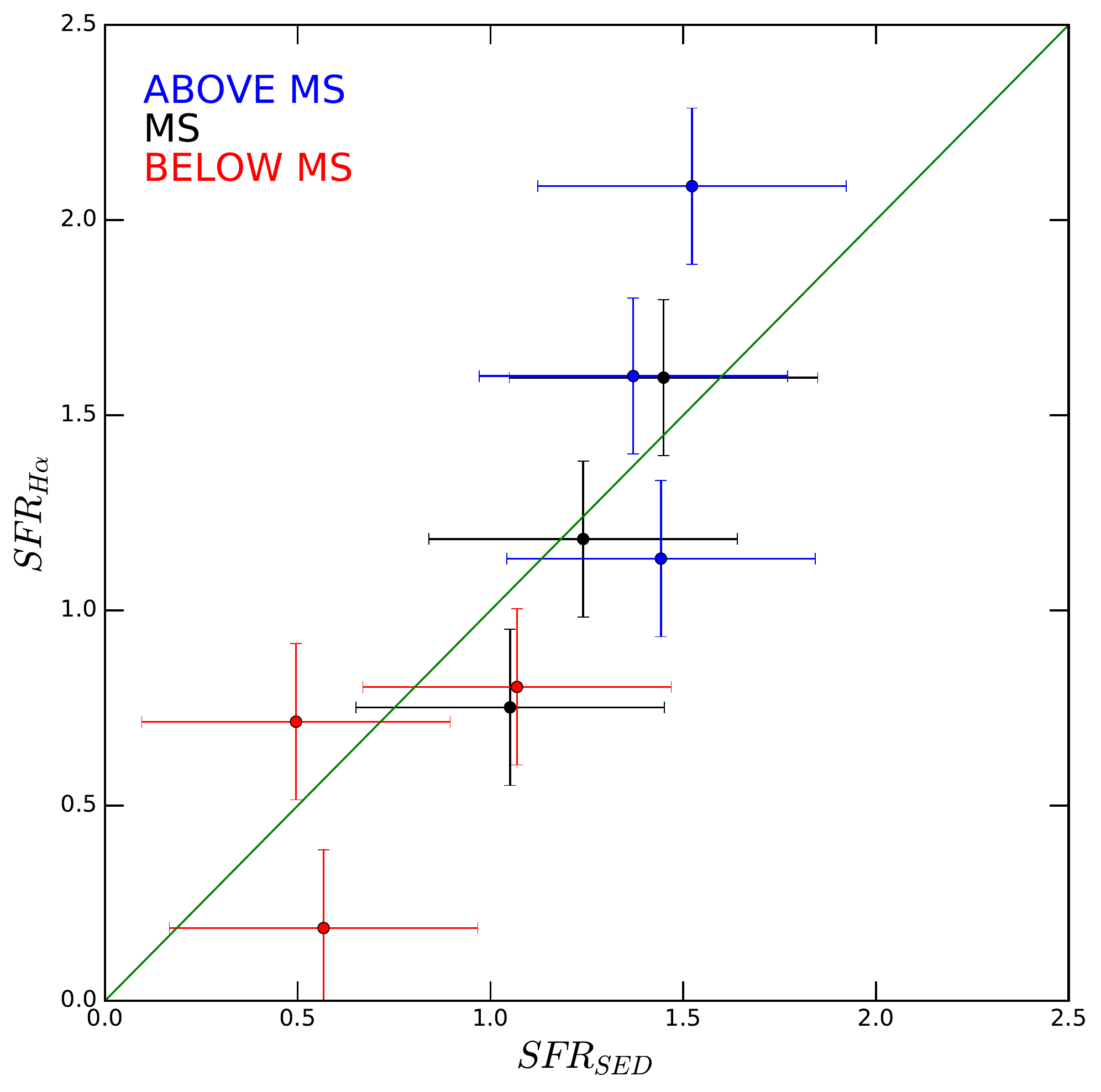}
\caption{SFR from H${\alpha}$ as obtained by integrating the SFR profiles of N16 in different bins of distance from the MS and stellar mass, compared to the SFR from SED fitting.The two are in agreement within the large errors associated to SFR from SED fitting.}
\label{fig:comp}
\end{figure}

Most of previous studies relate the enhancement and suppression of the galaxy central region to the stellar mass. 
\cite{2017arXiv171005049S} analyse the distribution of star formation activity in MaNGA galaxies and its dependence on integrated mass, structure, and environment. They propose the existence of two classes of galaxies, namely \textit{Centrally Suppressed} and \textit{Unsuppressed}, defined as galaxies that have sSFR in their disk at least 10 times larger than the core, and galaxies with flat sSFR profiles respectively. They find that there is a strong dependence of the classification of galaxies as \textit{Centrally Suppressed} and \textit{Unsuppressed} on stellar mass. This is consistent with our finding of more pronounced decrease of sSFR in the inner part of galaxies with increasing stellar mass. \textit{Centrally Suppressed} galaxies have up to $\sim$ 0.5 dex suppression in their sSFR at larger radii compared to \textit{Unsuppressed} galaxies, which is also the maximum difference seen in our data in the sSFR values at r $>$ 5 kpc for star-forming and quiescent galaxies. Finally, when studying the dependence of central suppression of star formation on galaxy morphology, Splinder et al. (2017) conclude that it is strongly related to the presence of a bulge, in agreement with our findings of increased S\'ersic index in galaxies with the deepest central depression in their sSFR profiles. 

\cite{2017arXiv171005034B} analyze the H$\alpha$ equivalent width profiles of MaNGA galaxies, and find that the typical sSFR profile of MS galaxies is flat at LogM$_{\star} <$ 10.5 M$_{\odot}$, while the sSFR of more massive MS galaxies shows a decrease in the central region. This result is partially in disagreement with our findings for the \textit{low-z} sample, as we observe that the sSFR profiles of MS galaxies is always smaller in the central 5 kpc than at larger radii even if the difference is only up to 0.5 dex. This discrepancy is most likely due to different selection criteria for MS galaxies in the two works. In fact, in \cite{2017arXiv171005034B} MS galaxies are not distinguishable from upper MS and SB galaxies, which might exhibit a centrally enhanced SFR profile. As a consequence, the median profile is flat and shows a decrease at inner radii only at larger stellar masses, when the depression in the SFR profile of MS galaxies becomes more significant. Our median SFR profiles as a function of galaxy distance from the MS are in qualitative agreement with the ones reported by \cite{2017arXiv171100915E}. In both studies, galaxies above the MS are characterized by a considerable enhancement of SFR surface density in the central regions, while passive galaxies show a depression in the inner part. The result of centrally-suppressed SFR profiles in passive galaxies is also in agreement with the study of \cite{2013IAUS..295..300D}, which analyze CALIFA galaxies to study the dependence of the SFR profile on morphological classification. They find that SFR profiles of disk-like galaxies increase towards the centre, while $S0$ galaxies and ellipticals have centrally suppressed profiles. This is compatible with our observations as galaxies whose SFR profile is suppressed at the centre also exhibit S\'ersic indexes comparable to bulge+disk systems, or pure spheroidal morphologies below the MS (see Fig. \ref{fig:sersic_mine}). 

As we discussed above, no evolution is seen between the \textit{low-z} and \textit{high-z} samples. The only marginal difference between the two samples is related to SB galaxies. In fact, we do not find many of them in the high redshift bin. Those few tend to exhibit a less significant central enhancement in the SFR profile with respect to the MS galaxies, as clearly visible, instead, in the low redshift bin. However, this could also be due to the low number statistics in the highest redshift bin, in which the SB region is not much populated in our dataset. This would suggest that the distribution of the SF activity in galaxies is driving the scatter of the MS in the same way at least up to z$\sim$1.2.


\cite{2016ApJ...828...27N} present the average surface SFR and sSFR profiles based on H${\alpha}$ luminosity in the 3D-HST sample. The subsample is, in particular, H${\alpha}$ flux-selected in the redshift range [0.7:1.5]. The authors find evidence of a "coherent" SFR distribution, whereas above the MS the SFR profile is on average enhanced at all radii, while below the MS, it is suppressed at all radii, but with a very similar shape. They also observe an enhancement of the SF activity in the central region towards the upper envelope of the MS, although less significant than in our results and limited to a small stellar mass range. The spatial region investigated by N16 is limited to the inner $\sim$ 6-8 kpc, while SED fitting gives us the possibility to explore a wider area, up to  $\sim$15-20 kpc. In addition, their below MS subsample includes  galaxies down to -0.8 dex from the MS and it does not sample sources in the quiescence region, as our SUB2 and QUIE subsamples. The two sets of SFR profiles are in agreement within the inner $\sim6-8$ kpc when galaxies are selected in a consistent way as a function of the distance from the MS. We find that the mean integrated H${\alpha}$-based SFR and the SED fitting-based SFR for the same class of galaxies agree well (see Fig.~\ref{fig:comp}) and are consistent within the errors. Those are obtained by integrating the H${\alpha}$-based SFR profiles of N16 and the HDUV profiles of this work for the same class of galaxies in the same redshift range. 
We notice instead a disagreement with the stellar mass distribution. While in N16 the stellar mass profiles of galaxies above, on, and below the MS are remarkably similar, in this work we find that the stellar mass distribution of galaxies well below the MS is always more compact and centrally peaked than MS counterparts, as also revealed by the study of the Sersic index and in agreement with previous studies (e.g. Whitaker et al. 2014 and references therein). As a consequence we obtain a sSFR profile more centrally suppressed than in N16 for galaxies below the MS. Furthermore, due to the H${\alpha}$ selection, in N16 the region below the MS is highly incomplete and potentially biased towards H${\alpha}$ bright sources of the subsample. This could lead to SFR profiles more similar to those of MS galaxies and less flat as observed in our work, where the completeness is much higher in the quiescence region.

Other recent works investigating the spatial distribution of SFR in galaxies are based on small samples for which high spatial resolution data is available. \cite{2015ApJ...802..101T} studied H${\alpha}$ derived SFR profiles in 22 galaxies at z $\sim$ 2 (resolution $\sim$ 1 kpc) and observed that the average profiles do not change with stellar mass (and are consistent with a disk-like profile). We also do not observe significant changes in the SFR profiles of MS galaxies at different stellar masses. More recently, \cite{2017arXiv170400733T} show dust, UV, and H$\alpha$ dust-corrected profiles for 10 galaxies at z $\sim$ 2 and found centrally suppressed sSFR profiles only in the most massive bin (M$_{\star}>10^{11}$M$_{\odot}$), while a flat sSFR profile is observed at M$_{\star}<10^{11}$M$_{\odot}$. In our sample, central suppression is always visible in the sSFR profiles of galaxies in the \textit{high-z} sample, but the suppression becomes more significant at increasing stellar masses. In the same work, the authors also argue that galaxies above the MS have sSFR characterised by a central enhancement, in agreement with what is suggested by the trends in our profiles.


\subsection{Stellar mass distribution across - along the MS}

While our work on the sSFR and SFR profiles is in clear agreement with previous results, the trends in stellar mass distribution across the MS and for galaxies with the largest SFRs present a challenge.  
In \cite{2017A&A...597A..97M} we study the the structure of z $\sim$ 0 SDSS galaxies using the B/T decomposition of \cite{2011ApJS..196...11S} and find that the MS of star-forming galaxies corresponds to the minimum of the B/T distribution at each stellar mass; in other words, galaxies in the upper envelope of the MS are characterised by a larger B/T with respect to MS counterparts. The result, while in agreement with the original work of \cite{2011ApJ...742...96W}, does not comply with the findings of this work where an opposite trend of decreasing S\'ersic index is observed when moving from the MS towards its upper envelope. This discrepancy could be ascribed to the fact that, since we employ the small inner region of the  GOODS sample, we are missing real starburst galaxies located more than one dex above the MS.These galaxies that we miss in our current sample could be pure spheroidals \citep{2017MNRAS.472.1401A}, the inclusion of which would increase the average S\'ersic index of galaxies above the MS. 
  
On the other hand, the morphological trend of increasing S\'ersic index along the MS is confirmed by different works in recent literature, both at low \citep{2014MNRAS.441..599B,2017A&A...597A..97M} and high \citep{2014ApJ...788...11L} redshift. Such trends, together with the more pronounced decrease of the sSFR of MS galaxies with increasing stellar mass, indicate that galaxies grow their bulge component while moving along the MS, and that star formation in massive MS galaxies takes place only in the disk, while the bulges remain passive, as argued in \cite{2015Sci...348..314T}

\subsection{Migration of galaxies across the MS and implications on quenching mechanism}

The shapes of the observed SFR and sSFR profiles are compatible with the compaction-depletion scenario of Tacchella et al. (2015b), based on the VELA simulations (Ceverino et al., 2014). These studies propose a scenario where star forming galaxies migrate across the MS due to the successive episodes of gas inflow and star formation ongoing in the central part. 
Here, cold gas flowing to the inner region of galaxies causes an overall compaction of the system, which in turn leads to a phase of high star formation activity. Following this, central depletion of gas due to star formation forces a galaxy below the MS.  As a result, sSFR profiles of galaxies are expected to be centrally enhanced (suppressed) above (below) the MS, which is in good agreement with our findings. \\
\\
The shape of the SFR and sSFR profiles is also a powerful tool to obtain information on the nature of the quenching mechanism. The coherent star formation picture presented in \cite{2016ApJ...828...27N} was used to conclude that the suppression of star formation at all radii, observed for galaxies below the MS, must be related to a quenching mechanism that acts not only in the central part of galaxies but also troughout the disk. {The central suppression observed in the sSFR profiles of galaxies below the MS is  compatible with the so-called $inside-out$ quenching mechanism, where the star formation halts first in the innermost radii and then in the outer regions of the disk, implying that disks are still forming stars when the central part of a galaxy is dominated by a passive bulge.} Such a quenching mechanism has been advocated frequently in recent studies. \cite{2016ApJ...816...87M} used HST/WFC3 and ACS observations to study UV rest frame colours of Milky Way progenitors in $0.5<z<3$. Despite the uncertainties in using UV colours as SFR tracers, they find that quenching must first occur in the central region, followed by the outskirts. \cite{2015Sci...348..314T} employ the centrally-suppressed sSFR profiles of massive MS galaxies to conclude that quenching must proceed in an $inside-out$ fashion, with timescales ranging from few hundred million years in the central regions to a few billion years at larger radii. In \cite{2016MNRAS.458..242T} the authors used cosmological simulations to extract a sample of 26 galaxies at $1<z<7$ and found that the onset of gas depletion occurs in the central 1 kpc and follows a phase of wet compaction and SFR peaked in the centre. The quenching hence also proceeds $inside-out$, implying that when the central region is passive star formation is still ongoing in an outer ring. A small sample of 10 star forming galaxies at z $\sim$ 2, observed as part of a SINS/zC-SINF AO program, is used in \cite{2017arXiv170400733T} to study dust and UV profiles. They find that only massive galaxies  (M$_{\star}>10^{11}$M$_{\odot}$) show sSFR profiles that are suppressed in the centre with respect to the outskirts and hence must be experiencing $inside-out$ quenching. More work needs to be done in order to understand the main driver behind the suppression of the central star formation, f.e. by SFHs of quiescent galaxies.


\subsection{Summary}

In this work, we exploit multi-wavelength, high-resolution HST imaging to study the spatially resolved properties of galaxies located in the central parts of the GOODS fields. We summarise our main findings as follows:
\begin{itemize}
\item the sSFR profiles of galaxies in the redshift range $0.2<z<1.2$ is centrally enhanced above the MS and centrally suppressed below the MS, with quiescent galaxies characterised by the deepest suppression. The sSFR in the outer region does not show systematic trends of enhancement or suppression above or below the MS. 
\item the S\'ersic index of MS galaxies increases with increasing stellar mass, indicating that bulges are growing when galaxies are still on the relation. Galaxies are also more bulge-dominated in the lower envelope of the MS, while galaxies in the upper envelope are more extended and have S\'ersic indexes that are always smaller than or comparable to their MS counterparts.
\end{itemize}

These findings suggest that galaxies above the MS could be in a evolutionary phase where cold gas is available in the central region, possibly due to minor mergers or disk instabilities, but no compaction has still taken place. Furthermore, our work indicates that \textit{inside-out} quenching is indeed taking place, as star formation is centrally-suppressed in galaxies below the MS, but still occurs in the outer regions. Follow up studies are warranted in order to: 1) increase the statistics, especially above the MS so as to cover the area at $\sim$1 dex from the relation, and 2) investigate the spatial distribution of cold gas so as to understand the reasons behind the growth of the bulge observed below the MS, e.g. the stabilisation of the disk against further collapse vs. removal of cold gas due to AGN activity.

\section*{Aknowledgements}
LM thanks Erica Nelson and Pieter van Dokkum for comments that greatly improved the analysis, and Sandro Tacchella for giving useful feedback on the manuscript. This work is based on observations made with the NASA/ESA Hubble Space Telescope, obtained from the Mikulski Archive for Space Telescopes (MAST) at the Space Telescope Science Institute. Support for this work was provided by NASA through grant HST-GO-13872 from the Space Telescope Science Institute, which is operated by AURA, Inc., under NASA contract NAS 5-26555. LM thanks Bhaskar Agarwal and Matias Blana for useful and stimulating discussion.

\bibliographystyle{mnras} 
\bibliography{laura} 

\begin{thebibliography}{}
\makeatletter
\relax
\def\mn@urlcharsother{\let\do\@makeother \do\$\do\&\do\#\do\^\do\_\do\%\do\~}
\def\mn@doi{\begingroup\mn@urlcharsother \@ifnextchar [ {\mn@doi@}
  {\mn@doi@[]}}
\def\mn@doi@[#1]#2{\def\@tempa{#1}\ifx\@tempa\@empty \href
  {http://dx.doi.org/#2} {doi:#2}\else \href {http://dx.doi.org/#2} {#1}\fi
  \endgroup}
\def\mn@eprint#1#2{\mn@eprint@#1:#2::\@nil}
\def\mn@eprint@arXiv#1{\href {http://arxiv.org/abs/#1} {{\tt arXiv:#1}}}
\def\mn@eprint@dblp#1{\href {http://dblp.uni-trier.de/rec/bibtex/#1.xml}
  {dblp:#1}}
\def\mn@eprint@#1:#2:#3:#4\@nil{\def\@tempa {#1}\def\@tempb {#2}\def\@tempc
  {#3}\ifx \@tempc \@empty \let \@tempc \@tempb \let \@tempb \@tempa \fi \ifx
  \@tempb \@empty \def\@tempb {arXiv}\fi \@ifundefined
  {mn@eprint@\@tempb}{\@tempb:\@tempc}{\expandafter \expandafter \csname
  mn@eprint@\@tempb\endcsname \expandafter{\@tempc}}}

\bibitem[\protect\citeauthoryear{Abramson, Kelson, Dressler, Poggianti,
  Gladders, Oemler  \& Vulcani}{Abramson et~al.}{2014}]{2014ApJ...785L..36A}
Abramson L.~E.,  Kelson D.~D.,  Dressler A.,  Poggianti B.,  Gladders M.~D.,
  Oemler A.~J.,   Vulcani B.,  2014, The Astrophysical Journal Letters, 785,
  L36

\bibitem[\protect\citeauthoryear{Allen, Driver, Graham, Cameron, Liske  \&
  De~Propris}{Allen et~al.}{2006}]{2006MNRAS.371....2A}
Allen P.~D.,  Driver S.~P.,  Graham A.~W.,  Cameron E.,  Liske J.,   De~Propris
  R.,  2006, arXiv.org, pp 2--18

\bibitem[\protect\citeauthoryear{Almaini et~al.,}{Almaini
  et~al.}{2017}]{2017MNRAS.472.1401A}
Almaini O.,  et~al., 2017, Monthly Notices of the Royal Astronomical Society,
  472, 1401

\bibitem[\protect\citeauthoryear{Arnouts, Cristiani, Moscardini, Matarrese,
  Lucchin, Fontana  \& Giallongo}{Arnouts et~al.}{1999}]{1999MNRAS.310..540A}
Arnouts S.,  Cristiani S.,  Moscardini L.,  Matarrese S.,  Lucchin F.,  Fontana
  A.,   Giallongo E.,  1999, Monthly Notices, 310, 540

\bibitem[\protect\citeauthoryear{Balogh, Baldry, Nichol, Miller, Bower  \&
  Glazebrook}{Balogh et~al.}{2004}]{2004ApJ...615L.101B}
Balogh M.~L.,  Baldry I.~K.,  Nichol R.,  Miller C.,  Bower R.,   Glazebrook
  K.,  2004, The Astrophysical Journal, 615, L101

\bibitem[\protect\citeauthoryear{Belfiore et~al.,}{Belfiore
  et~al.}{2017}]{2017arXiv171005034B}
Belfiore F.,  et~al., 2017, arXiv.org, p. arXiv:1710.05034

\bibitem[\protect\citeauthoryear{Bell et~al.,}{Bell
  et~al.}{2005}]{2005ApJ...625...23B}
Bell E.~F.,  et~al., 2005, The Astrophysical Journal, 625, 23

\bibitem[\protect\citeauthoryear{Bernhard, Bethermin, Sargent, Buat, Mullaney,
  Pannella, Heinis  \& Daddi}{Bernhard et~al.}{2014}]{2014MNRAS.442..509B}
Bernhard E.,  Bethermin M.,  Sargent M.,  Buat V.,  Mullaney J.~R.,  Pannella
  M.,  Heinis S.,   Daddi E.,  2014, Monthly Notices of the Royal Astronomical
  Society, 442, 509

\bibitem[\protect\citeauthoryear{Blanton \& Moustakas}{Blanton \&
  Moustakas}{2009}]{2009ARA&A..47..159B}
Blanton M.~R.,  Moustakas J.,  2009, Annual Review of Astronomy and
  Astrophysics, 47, 159

\bibitem[\protect\citeauthoryear{Blanton, Eisenstein, Hogg, Schlegel  \&
  Brinkmann}{Blanton et~al.}{2005}]{2005ApJ...629..143B}
Blanton M.~R.,  Eisenstein D.,  Hogg D.~W.,  Schlegel D.~J.,   Brinkmann J.,
  2005, The Astrophysical Journal, 629, 143

\bibitem[\protect\citeauthoryear{Bluck, Mendel, Ellison, Moreno, Simard, Patton
   \& Starkenburg}{Bluck et~al.}{2014}]{2014MNRAS.441..599B}
Bluck A. F.~L.,  Mendel J.~T.,  Ellison S.~L.,  Moreno J.,  Simard L.,  Patton
  D.~R.,   Starkenburg E.,  2014, Monthly Notices of the Royal Astronomical
  Society, 441, 599

\bibitem[\protect\citeauthoryear{Brusa et~al.,}{Brusa
  et~al.}{2010}]{2010ApJ...716..348B}
Brusa M.,  et~al., 2010, The Astrophysical Journal, 716, 348

\bibitem[\protect\citeauthoryear{Bruzual \& Charlot}{Bruzual \&
  Charlot}{2003}]{2003MNRAS.344.1000B}
Bruzual G.,  Charlot S.,  2003, Monthly Notices of the Royal Astronomical
  Society, 344, 1000

\bibitem[\protect\citeauthoryear{Calzetti, Armus, Bohlin, Kinney, Koornneef  \&
  Storchi-Bergmann}{Calzetti et~al.}{2000}]{2000ApJ...533..682C}
Calzetti D.,  Armus L.,  Bohlin R.~C.,  Kinney A.~L.,  Koornneef J.,
  Storchi-Bergmann T.,  2000, The Astrophysical Journal, 533, 682

\bibitem[\protect\citeauthoryear{Chabrier}{Chabrier}{2003}]{2003PASP..115..763C}
Chabrier G.,  2003, The Publications of the Astronomical Society of the
  Pacific, 115, 763

\bibitem[\protect\citeauthoryear{Cheung et~al.,}{Cheung
  et~al.}{2012}]{2012ApJ...760..131C}
Cheung E.,  et~al., 2012, The Astrophysical Journal, 760, 131

\bibitem[\protect\citeauthoryear{Cibinel et~al.,}{Cibinel
  et~al.}{2015}]{2015ApJ...805..181C}
Cibinel A.,  et~al., 2015, The Astrophysical Journal, 805, 181

\bibitem[\protect\citeauthoryear{Daddi et~al.,}{Daddi
  et~al.}{2007}]{2007ApJ...670..156D}
Daddi E.,  et~al., 2007, The Astrophysical Journal, 670, 156

\bibitem[\protect\citeauthoryear{Delgado et~al.,}{Delgado
  et~al.}{2013}]{2013IAUS..295..300D}
Delgado R.~G.,  et~al., 2013, The Intriguing Life of Massive Galaxies, 295, 300

\bibitem[\protect\citeauthoryear{Dressler}{Dressler}{1980}]{1980ApJ...236..351D}
Dressler A.,  1980, Astrophysical Journal, 236, 351

\bibitem[\protect\citeauthoryear{Drory et~al.,}{Drory
  et~al.}{2009}]{Drory:2009cp}
Drory N.,  et~al., 2009, arXiv.org, pp 1595--1609

\bibitem[\protect\citeauthoryear{Elbaz et~al.,}{Elbaz
  et~al.}{2007}]{2007A&A...468...33E}
Elbaz D.,  et~al., 2007, Astronomy {\&} Astrophysics, 468, 33

\bibitem[\protect\citeauthoryear{Elbaz et~al.,}{Elbaz
  et~al.}{2011}]{2011A&A...533A.119E}
Elbaz D.,  et~al., 2011, Astronomy {\&} Astrophysics, 533, A119

\bibitem[\protect\citeauthoryear{Ellison, Sanchez, Ibarra-Medel, Antonio,
  Mendel  \& Barrera-Ballesteros}{Ellison et~al.}{2017}]{2017arXiv171100915E}
Ellison S.~L.,  Sanchez S.~F.,  Ibarra-Medel H.,  Antonio B.,  Mendel J.~T.,
  Barrera-Ballesteros J.,  2017, arXiv.org, p. arXiv:1711.00915

\bibitem[\protect\citeauthoryear{Erfanianfar et~al.,}{Erfanianfar
  et~al.}{2016}]{2016MNRAS.455.2839E}
Erfanianfar G.,  et~al., 2016, Monthly Notices of the Royal Astronomical
  Society, 455, 2839

\bibitem[\protect\citeauthoryear{Giavalisco et~al.,}{Giavalisco
  et~al.}{2004}]{2004ApJ...600L..93G}
Giavalisco M.,  et~al., 2004, The Astrophysical Journal, 600, L93

\bibitem[\protect\citeauthoryear{Goddard et~al.,}{Goddard
  et~al.}{2017}]{2017MNRAS.465..688G}
Goddard D.,  et~al., 2017, Monthly Notices of the Royal Astronomical Society,
  465, 688

\bibitem[\protect\citeauthoryear{Grogin et~al.,}{Grogin
  et~al.}{2011}]{2011ApJS..197...35G}
Grogin N.~A.,  et~al., 2011, The Astrophysical Journal Supplement, 197, 35

\bibitem[\protect\citeauthoryear{Guo, Zheng, Wang  \& Fu}{Guo
  et~al.}{2015}]{Guo:2015kp}
Guo K.,  Zheng X.~Z.,  Wang T.,   Fu H.,  2015, arXiv.org, p.~L49

\bibitem[\protect\citeauthoryear{Ilbert et~al.,}{Ilbert
  et~al.}{2006}]{2006A&A...457..841I}
Ilbert O.,  et~al., 2006, Astronomy {\&} Astrophysics, 457, 841

\bibitem[\protect\citeauthoryear{Ilbert et~al.,}{Ilbert
  et~al.}{2009}]{2010ApJ...709..644I}
Ilbert O.,  et~al., 2009, arXiv.org, pp 644--663

\bibitem[\protect\citeauthoryear{Karim et~al.,}{Karim
  et~al.}{2011}]{2011ApJ...730...61K}
Karim A.,  et~al., 2011, The Astrophysical Journal, 730, 61

\bibitem[\protect\citeauthoryear{Kauffmann et~al.,}{Kauffmann
  et~al.}{2003}]{2003MNRAS.341...54K}
Kauffmann G.,  et~al., 2003, Monthly Notice of the Royal Astronomical Society,
  341, 54

\bibitem[\protect\citeauthoryear{Koekemoer et~al.,}{Koekemoer
  et~al.}{2011}]{2011ApJS..197...36K}
Koekemoer A.~M.,  et~al., 2011, The Astrophysical Journal Supplement, 197, 36

\bibitem[\protect\citeauthoryear{Kriek, van Dokkum, Franx, Illingworth  \&
  Magee}{Kriek et~al.}{2009}]{2009ApJ...705L..71K}
Kriek M.,  van Dokkum P.~G.,  Franx M.,  Illingworth G.~D.,   Magee D.~K.,
  2009, The Astrophysical Journal Letters, 705, L71

\bibitem[\protect\citeauthoryear{Lang et~al.,}{Lang
  et~al.}{2014}]{2014ApJ...788...11L}
Lang P.,  et~al., 2014, The Astrophysical Journal, 788, 11

\bibitem[\protect\citeauthoryear{Lin et~al.,}{Lin
  et~al.}{2017}]{2017arXiv171008610L}
Lin L.,  et~al., 2017, arXiv.org, p. arXiv:1710.08610

\bibitem[\protect\citeauthoryear{Lutz et~al.,}{Lutz
  et~al.}{2011}]{2011A&A...532A..90L}
Lutz D.,  et~al., 2011, Astronomy {\&} Astrophysics, 532, A90

\bibitem[\protect\citeauthoryear{Magnelli et~al.,}{Magnelli
  et~al.}{2013}]{2013A&A...553A.132M}
Magnelli B.,  et~al., 2013, Astronomy {\&} Astrophysics, 553, A132

\bibitem[\protect\citeauthoryear{Magnelli et~al.,}{Magnelli
  et~al.}{2014}]{2014A&A...561A..86M}
Magnelli B.,  et~al., 2014, Astronomy {\&} Astrophysics, 561, A86

\bibitem[\protect\citeauthoryear{Maragkoudakis, Zezas, Ashby  \&
  Willner}{Maragkoudakis et~al.}{2017}]{2017MNRAS.466.1192M}
Maragkoudakis A.,  Zezas A.,  Ashby M. L.~N.,   Willner S.~P.,  2017, Monthly
  Notices of the Royal Astronomical Society, 466, 1192

\bibitem[\protect\citeauthoryear{Martig, Bournaud, Teyssier  \& Dekel}{Martig
  et~al.}{2009}]{2009ApJ...707..250M}
Martig M.,  Bournaud F.,  Teyssier R.,   Dekel A.,  2009, The Astrophysical
  Journal, 707, 250

\bibitem[\protect\citeauthoryear{Meurer, Heckman  \& Calzetti}{Meurer
  et~al.}{1999}]{1999ApJ...521...64M}
Meurer G.~R.,  Heckman T.~M.,   Calzetti D.,  1999, The Astrophysical Journal,
  521, 64

\bibitem[\protect\citeauthoryear{Momcheva et~al.,}{Momcheva
  et~al.}{2016}]{2016ApJS..225...27M}
Momcheva I.~G.,  et~al., 2016, The Astrophysical Journal Supplement Series,
  225, 27

\bibitem[\protect\citeauthoryear{Morgan \& Mayall}{Morgan \&
  Mayall}{1957}]{1957PASP...69..291M}
Morgan W.~W.,  Mayall N.~U.,  1957, Publications of the Astronomical Society of
  the Pacific, 69, 291

\bibitem[\protect\citeauthoryear{Morishita \& Ichikawa}{Morishita \&
  Ichikawa}{2016}]{2016ApJ...816...87M}
Morishita T.,  Ichikawa T.,  2016, The Astrophysical Journal, 816, 87

\bibitem[\protect\citeauthoryear{Morselli, Popesso, Erfanianfar  \&
  Concas}{Morselli et~al.}{2017}]{2017A&A...597A..97M}
Morselli L.,  Popesso P.,  Erfanianfar G.,   Concas A.,  2017, Astronomy {\&}
  Astrophysics, 597, A97

\bibitem[\protect\citeauthoryear{Muzzin et~al.,}{Muzzin
  et~al.}{2013}]{2013ApJ...777...18M}
Muzzin A.,  et~al., 2013, The Astrophysical Journal, 777, 18

\bibitem[\protect\citeauthoryear{Nelson et~al.,}{Nelson
  et~al.}{2016}]{2016ApJ...828...27N}
Nelson E.~J.,  et~al., 2016, The Astrophysical Journal, 828, 27

\bibitem[\protect\citeauthoryear{Noeske et~al.,}{Noeske
  et~al.}{2007}]{2007ApJ...660L..47N}
Noeske K.~G.,  et~al., 2007, The Astrophysical Journal, 660, L47

\bibitem[\protect\citeauthoryear{Oesch et~al.,}{Oesch
  et~al.}{2018}]{2018arXiv180601853O}
Oesch P.~A.,  et~al., 2018, arXiv.org, p. arXiv:1806.01853

\bibitem[\protect\citeauthoryear{Overzier et~al.,}{Overzier
  et~al.}{2011}]{2011ApJ...726L...7O}
Overzier R.~A.,  et~al., 2011, The Astrophysical Journal Letters, 726, L7

\bibitem[\protect\citeauthoryear{Pannella, Hopp, Saglia, Bender, Drory,
  Salvato, Gabasch  \& Feulner}{Pannella et~al.}{2006}]{2006ApJ...639L...1P}
Pannella M.,  Hopp U.,  Saglia R.~P.,  Bender R.,  Drory N.,  Salvato M.,
  Gabasch A.,   Feulner G.,  2006, arXiv.org, pp L1--L4

\bibitem[\protect\citeauthoryear{Pannella et~al.,}{Pannella
  et~al.}{2009}]{2009ApJ...698L.116P}
Pannella M.,  et~al., 2009, arXiv.org, pp L116--L120

\bibitem[\protect\citeauthoryear{Peng}{Peng}{2010}]{2010AAS...21522909P}
Peng C.,  2010, American Astronomical Society, 215, 229.09

\bibitem[\protect\citeauthoryear{Peng et~al.,}{Peng
  et~al.}{2010}]{2010ApJ...721..193P}
Peng Y.-j.,  et~al., 2010, The Astrophysical Journal, 721, 193

\bibitem[\protect\citeauthoryear{Popesso et~al.,}{Popesso
  et~al.}{2012}]{2012A&A...537A..58P}
Popesso P.,  et~al., 2012, Astronomy {\&} Astrophysics, 537, A58

\bibitem[\protect\citeauthoryear{Pozzetti et~al.,}{Pozzetti
  et~al.}{2009}]{Pozzetti:2009gw}
Pozzetti L.,  et~al., 2009, arXiv.org, p.~A13

\bibitem[\protect\citeauthoryear{Renzini \& Peng}{Renzini \&
  Peng}{2015}]{2015ApJ...801L..29R}
Renzini A.,  Peng Y.-j.,  2015, The Astrophysical Journal Letters, 801, L29

\bibitem[\protect\citeauthoryear{Robert C~Kennicutt}{Robert
  C~Kennicutt}{1998}]{1998ARA&A..36..189K}
Robert C~Kennicutt J.,  1998, arXiv.org, pp 189--232

\bibitem[\protect\citeauthoryear{Rodighiero et~al.,}{Rodighiero
  et~al.}{2010}]{2010A&A...518L..25R}
Rodighiero G.,  et~al., 2010, Astronomy {\&} Astrophysics, 518, L25

\bibitem[\protect\citeauthoryear{Schreiber et~al.,}{Schreiber
  et~al.}{2015}]{2015A&A...575A..74S}
Schreiber C.,  et~al., 2015, Astronomy {\&} Astrophysics, 575, A74

\bibitem[\protect\citeauthoryear{Shao et~al.,}{Shao
  et~al.}{2010}]{2010A&A...518L..26S}
Shao L.,  et~al., 2010, Astronomy {\&} Astrophysics, 518, L26

\bibitem[\protect\citeauthoryear{Simard, Mendel, Patton, Ellison  \&
  McConnachie}{Simard et~al.}{2011}]{2011ApJS..196...11S}
Simard L.,  Mendel J.~T.,  Patton D.~R.,  Ellison S.~L.,   McConnachie A.~W.,
  2011, The Astrophysical Journal Supplement, 196, 11

\bibitem[\protect\citeauthoryear{Skelton et~al.,}{Skelton
  et~al.}{2014}]{2014ApJS..214...24S}
Skelton R.~E.,  et~al., 2014, The Astrophysical Journal Supplement Series, 214,
  24

\bibitem[\protect\citeauthoryear{Speagle, Steinhardt, Capak  \&
  Silverman}{Speagle et~al.}{2014}]{2014ApJS..214...15S}
Speagle J.~S.,  Steinhardt C.~L.,  Capak P.~L.,   Silverman J.~D.,  2014,
  arXiv.org, p.~15

\bibitem[\protect\citeauthoryear{Spindler et~al.,}{Spindler
  et~al.}{2017}]{2017arXiv171005049S}
Spindler A.,  et~al., 2017, arXiv.org, p. arXiv:1710.05049

\bibitem[\protect\citeauthoryear{Steinhardt, Yurk  \& Capak}{Steinhardt
  et~al.}{2017}]{2017MNRAS.468..849S}
Steinhardt C.~L.,  Yurk D.,   Capak P.,  2017, Monthly Notices of the Royal
  Astronomical Society, 468, 849

\bibitem[\protect\citeauthoryear{Szomoru, Franx, van Dokkum, Trenti,
  Illingworth, Labb{\'e}  \& Oesch}{Szomoru et~al.}{2013}]{2013ApJ...763...73S}
Szomoru D.,  Franx M.,  van Dokkum P.~G.,  Trenti M.,  Illingworth G.~D.,
  Labb{\'e} I.,   Oesch P.,  2013, The Astrophysical Journal, 763, 73

\bibitem[\protect\citeauthoryear{Tacchella et~al.,}{Tacchella
  et~al.}{2015a}]{2015Sci...348..314T}
Tacchella S.,  et~al., 2015a, Science, 348, 314

\bibitem[\protect\citeauthoryear{Tacchella et~al.,}{Tacchella
  et~al.}{2015b}]{2015ApJ...802..101T}
Tacchella S.,  et~al., 2015b, The Astrophysical Journal, 802, 101

\bibitem[\protect\citeauthoryear{Tacchella, Dekel, Carollo, Ceverino, DeGraf,
  Lapiner, Mandelker  \& Primack}{Tacchella et~al.}{2016}]{2016MNRAS.458..242T}
Tacchella S.,  Dekel A.,  Carollo C.~M.,  Ceverino D.,  DeGraf C.,  Lapiner S.,
   Mandelker N.,   Primack J.~R.,  2016, Monthly Notices of the Royal
  Astronomical Society, 458, 242

\bibitem[\protect\citeauthoryear{Tacchella et~al.,}{Tacchella
  et~al.}{2017a}]{2017arXiv170400733T}
Tacchella S.,  et~al., 2017a, arXiv.org, p. arXiv:1704.00733

\bibitem[\protect\citeauthoryear{Tacchella et~al.,}{Tacchella
  et~al.}{2017b}]{2018ApJ...859...56T}
Tacchella S.,  et~al., 2017b, arXiv.org, p.~56

\bibitem[\protect\citeauthoryear{Tal, van Dokkum, Franx, Leja, Wake  \&
  Whitaker}{Tal et~al.}{2013}]{2013ApJ...769...31T}
Tal T.,  van Dokkum P.~G.,  Franx M.,  Leja J.,  Wake D.~A.,   Whitaker K.~E.,
  2013, The Astrophysical Journal, 769, 31

\bibitem[\protect\citeauthoryear{Whitaker, van Dokkum, Brammer  \&
  Franx}{Whitaker et~al.}{2012}]{2012ApJ...754L..29W}
Whitaker K.~E.,  van Dokkum P.~G.,  Brammer G.,   Franx M.,  2012, The
  Astrophysical Journal Letters, 754, L29

\bibitem[\protect\citeauthoryear{Whitaker et~al.,}{Whitaker
  et~al.}{2014}]{2014ApJ...795..104W}
Whitaker K.~E.,  et~al., 2014, The Astrophysical Journal, 795, 104

\bibitem[\protect\citeauthoryear{Wuyts et~al.,}{Wuyts
  et~al.}{2011}]{2011ApJ...742...96W}
Wuyts S.,  et~al., 2011, The Astrophysical Journal, 742, 96

\bibitem[\protect\citeauthoryear{Zibetti}{Zibetti}{2009}]{2009arXiv0911.4956Z}
Zibetti S.,  2009, arXiv.org, p. arXiv:0911.4956

\bibitem[\protect\citeauthoryear{Ziparo et~al.,}{Ziparo
  et~al.}{2013}]{2013MNRAS.434.3089Z}
Ziparo F.,  et~al., 2013, Monthly Notices of the Royal Astronomical Society,
  434, 3089

\bibitem[\protect\citeauthoryear{Zolotov et~al.,}{Zolotov
  et~al.}{2015}]{2015MNRAS.450.2327Z}
Zolotov A.,  et~al., 2015, Monthly Notices of the Royal Astronomical Society,
  450, 2327

\bibitem[\protect\citeauthoryear{van~der Wel et~al.,}{van~der Wel
  et~al.}{2012}]{2012ApJS..203...24V}
van~der Wel A.,  et~al., 2012, The Astrophysical Journal Supplement, 203, 24

\makeatother
\end{thebibliography}

\begin{appendix}

\section{Radial profiles normalised for effective radii}
\label{app}

\begin{figure*}
\centering
\includegraphics[width=0.85\textwidth]{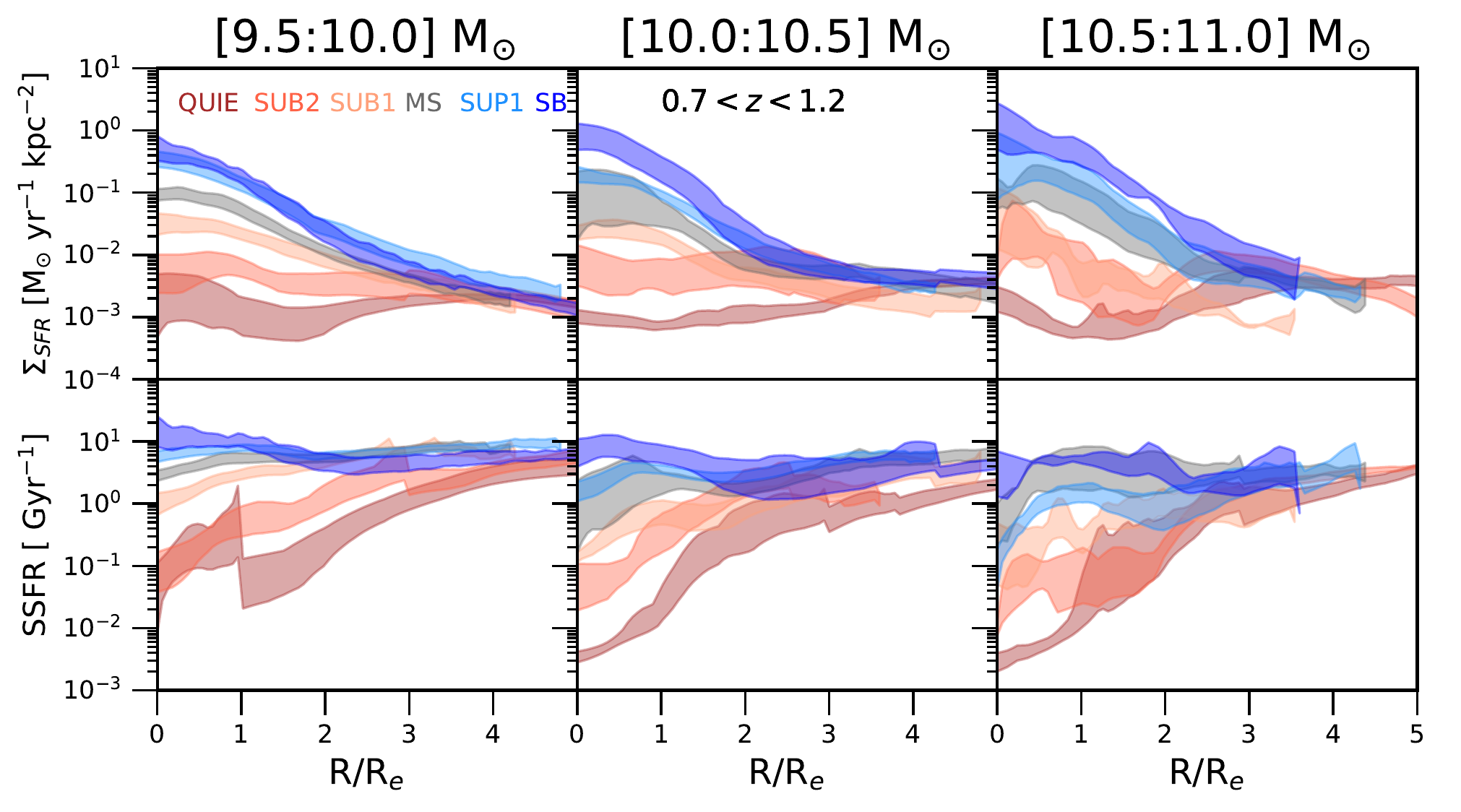}
\caption{Median SFR (first row), stellar mass (second row), and sSFR (third row) surface densities as a function of distance from the centre  (in R/R$_e$) for galaxies in the redshift range [0.2:0.7]. The three columns indicate galaxies with different stellar masses: 9.5$<$Log(M$_{\star}$/M$_{\odot}$)$<$10.0 in the first, 10.0$<$Log(M$_{\star}$/M$_{\odot}$)$<$10.5 in the second, and 10.5$<$Log(M$_{\star}$/M$_{\odot}$)$<$11.0 in the third column, respectively. Different colours mark different bins of distance from the MS, as defined in Fig.~\ref{fig:bins}. In the second row, the average S\'ersic indexes of galaxies in the 6 bins are written at the top of the panel, with colours corresponding to the bins. We observe than the trend of decreasing sSFR surface density in the central region of galaxies with increasing distance below the MS is also visible when considering the distance from the centre normalised by the mass-weighted effective radii.}
\label{fig:all_lowz_re}
\end{figure*}

\begin{figure*}
\centering
\includegraphics[width=0.85\textwidth]{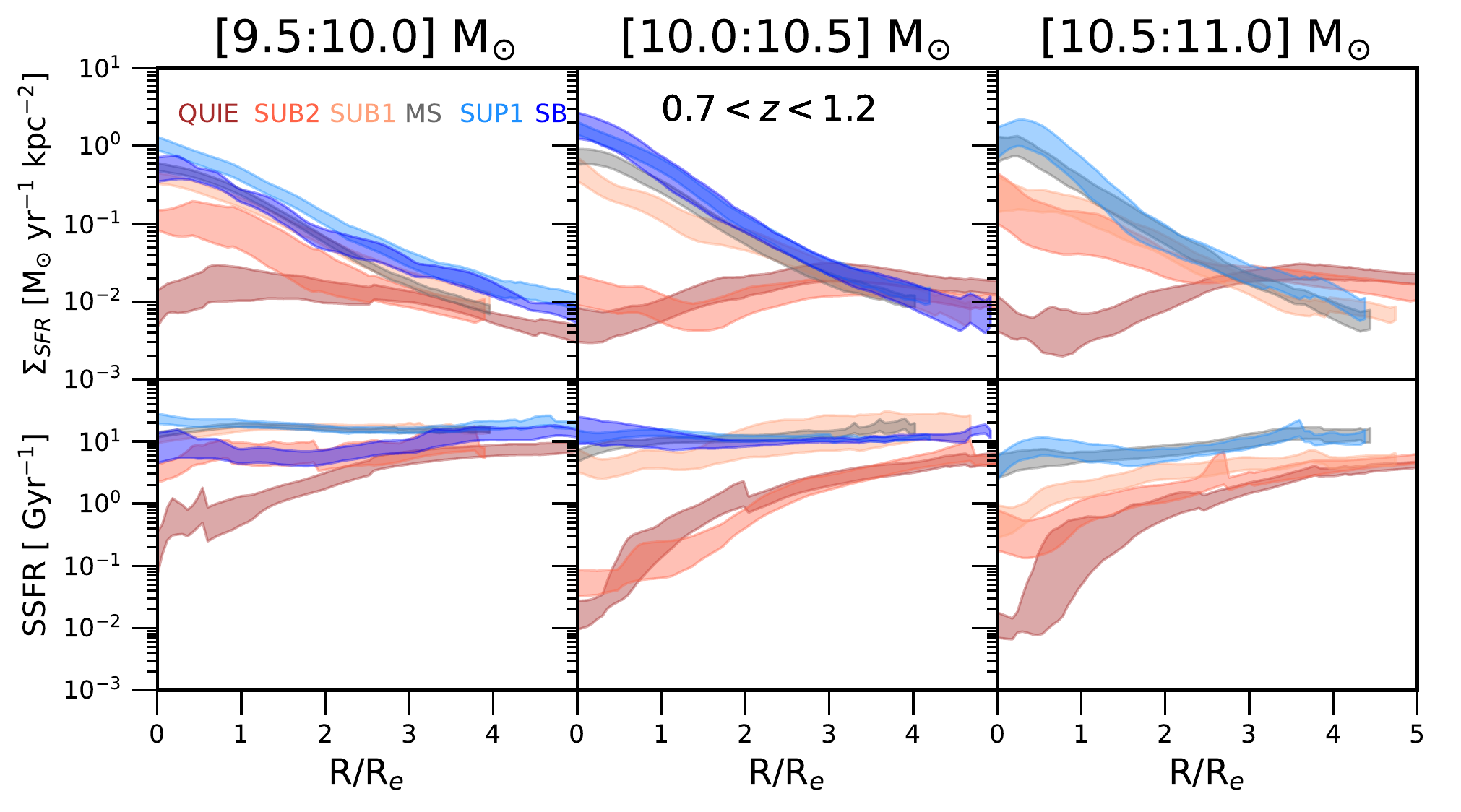}
\caption[SFR, M$_{\star}$ and sSFR profiles (in R/R$_e$) of \textit{high-z} sample.]{Same as Fig.\ref{fig:all_lowz_re}, but for galaxies in the redshift range [0.7:1.2]. Also in this redshift bin the sSFR surface density in the central region of galaxies decreases significantly when moving from the MS towards lower SFRs.}
\label{fig:high_z_re}
\end{figure*}

Here we show the SFR and sSFR surface densities median profiles, where the distance from the centre of each galaxy has been normalised to the mass-weighted effective radius. Fig.~\ref{fig:all_lowz_re} shows the result for the \textit{low-z} sample, and Fig.~\ref{fig:high_z_re} for the \textit{high-z} one, respectively. The trends in the SFR and sSFR surface density, computed as a function of the physical distance from the galaxy centre and analysed in Sec.~\ref{sec:results}, are still observed now that the distance is normalised for the mass-weighted effective radii. Galaxies above the MS and galaxies below it have SFR surface densities that differ of up to $\sim$ 2.5 order of magnitudes within 1 R$_\text{e}$, while in the outer regions the SFR surface densities of galaxies are comparable and independent on their distance from the relation.  Analogously, differences up to $\sim$ 2.5 orders of magnitude are observed within one R$_\text{e}$ between the sSFR surface density of galaxies in the QUIE and SUP1/SB samples, and such a difference become less significant at larger radii. As discussed in Sec.~\ref{sec:discussion} we consider such trends to be evidence of \textit{inside-out} quenching.

 \end{appendix}
 \end{document}